%% file: main.tex
\documentclass[journal]{IEEEtran}

\usepackage[acronym]{glossaries}
\usepackage[dvipsnames]{xcolor}

\usepackage{tikz}
\usepackage{tikz-3dplot}
\usetikzlibrary{shapes.multipart}
\usetikzlibrary{positioning}
\usetikzlibrary{arrows}
\usetikzlibrary{shapes.multipart}
\usetikzlibrary{shapes.geometric,shapes.arrows,decorations.pathmorphing}
\usetikzlibrary{matrix,chains,scopes,positioning,arrows,fit}
\usetikzlibrary{shapes.multipart}
\usetikzlibrary{calc}
\usetikzlibrary{decorations.markings}
\usetikzlibrary{shapes.geometric}
\usetikzlibrary{positioning}
\usepackage{pgfplots}
\pgfplotsset{compat=1.18}
\usepgfplotslibrary{dateplot, statistics}

\tikzset{
    msgarrow/.style={
        dashed, color = gray!60!black
        },
}

\usepackage{bbm}

\usepackage{amsmath}
\usepackage{amssymb}

\usepackage{algorithm}
\usepackage{algpseudocode}

\usepackage{subcaption}
\usepackage{cite}

\usepackage[inline]{enumitem}

\newcommand{\vek}[1]{\boldsymbol{#1}}

\newcommand{\imag}{\mathrm{j}}
\newcommand{\e}{\mathrm{e}}

\newcommand{\tsym}{{T}}

\newcommand{\nsym}{\ensuremath{n}}
\newcommand{\nstages}{\ensuremath{S}}
\newcommand{\nchannels}{\ensuremath{C}}
\newcommand{\nrings}{{\ensuremath{n_r}}}
\newcommand{\nsequence}{\ensuremath{{N_\text{seq}}}}
\newcommand{\ntraining}{\ensuremath{{N_\text{train}}}}
\newcommand{\incmean}{\ensuremath{\mu_\delta}}
\newcommand{\incvar}{\ensuremath{\sigma_\delta^2}}
\newcommand{\incstd}{\ensuremath{\sigma_\delta}}
\newcommand{\thetavar}{\ensuremath{\sigma_\theta^2}}

\newcommand{\ptx}{{P}}
\newcommand{\modpi}[1]{\ensuremath{m\left(#1\right)}}
\newcommand{\const}{\ensuremath{\mathrm{const.}}}
\newcommand{\kurtosis}{\ensuremath{Q}}

\newcommand{\appref}[1]{Appendix \ref{#1}}
\newcommand{\muF}{\overrightarrow{\mu}}
\newcommand{\muB}{\overleftarrow{\mu}}
\newcommand{\etaF}{\overrightarrow{\eta}}
\newcommand{\etaB}{\overleftarrow{\eta}}
\newcommand{\etaFvec}{\ensuremath{\overrightarrow{\vek{\eta}}}}
\newcommand{\etaBvec}{\ensuremath{\overleftarrow{\vek{\eta}}}}
\newcommand{\sigmaF}{\ensuremath{\overrightarrow{\sigma}}}
\newcommand{\sigmaB}{\ensuremath{\overleftarrow{\sigma}}}
\newcommand{\normpdf}[3]{{\ensuremath{\mathcal{N}\left(#1;#2,#3\right)}}}
\newcommand{\cscgnormpdf}[3]{{\ensuremath{\mathcal{N}_\mathbb{C}\left(#1;#2,#3\right)}}}
\newcommand{\cnormpdf}[4]{{\ensuremath{\mathcal{N}_\mathbb{C}\left(#1;#2,#3,#4\right)}}}

\newcommand{\normF}[1]{\ensuremath{\frac{1}{\overrightarrow{c}_{#1}}}}
\newcommand{\normB}[1]{\ensuremath{\frac{1}{\overleftarrow{c}_{#1}}}}
\newcommand{\anglebr}[1]{{\ensuremath{\angle\left(#1 \right)}}}
\newcommand{\anglenobr}[1]{{\ensuremath{\angle #1 }}}
\renewcommand{\mod}{{\ensuremath{\ \mathrm{mod} \ }}}

\definecolor{matlabBlue}{rgb}{0.00000,0.44700,0.74100}%
\definecolor{matlabRed}{rgb}{0.85000,0.32500,0.09800}%
\definecolor{matlab1}{rgb}{0.00000,0.44700,0.74100}%
\definecolor{matlab2}{rgb}{0.85000,0.32500,0.09800}%
\definecolor{matlab3}{rgb}{0.92900,0.69400,0.12500}%
\definecolor{matlab4}{rgb}{0.49400,0.18400,0.55600}%
\definecolor{matlab5}{rgb}{0.46600,0.67400,0.18800}%
\definecolor{matlab6}{rgb}{0.30100,0.74500,0.93300}%
\definecolor{matlab7}{rgb}{0.63500,0.07800,0.18400}%

\newcommand{\figref}[1]{Fig.~\ref{#1}}
\newcommand{\tabref}[1]{Table~\ref{#1}}
\newcommand{\secref}[1]{Sec.~\ref{#1}}
\usepackage[detect-all]{siunitx}
\DeclareSIUnit\bpcu{\gls{bpcu}}

\loadglsentries{glossary}

\begin{document}
\title{Information Rates of Successive Interference Cancellation for Optical Fiber}

\author{Alex~Jäger and~Gerhard~Kramer,~\IEEEmembership{Fellow,~IEEE}%

\thanks{Alex Jäger and Gerhard Kramer are with the Institute of Communications Engineering, Department of Computer Engineering, School of Computation, Information and Technology, Technical University of Munich, 80333 Munich, Germany (e-mail: alex.jaeger@tum.de; gerhard.kramer@tum.de).}%
}

\maketitle

\begin{tikzpicture}[remember picture,overlay]
\node[yshift = -1cm] at (current page.north){\parbox{\textwidth}{\footnotesize This work has been submitted to the IEEE for possible publication. Copyright may be transferred without notice, after which this version may no longer be accessible.}};
\end{tikzpicture}

\begin{abstract}
\Gls{jdd} achieves rates based on information theory but is too complex to implement for many channels with memory or nonlinearities. 
\Gls{sic} at the receiver, combined with multistage encoding at the transmitter, is a method that lets one use coded modulation for memoryless channels to approach \gls{jdd} rates. A \gls{sic}-based receiver is presented to compensate for inter-channel interference in long-haul optical fiber links. Simulations for \SI{1000}{\kilo\meter} of standard single-mode fiber with ideal distributed Raman amplification, single-polarization transmission, and \gls{cscg} modulation show that \gls{sic} attains the \glspl{air} of \gls{jdd} using surrogate channel models with \gls{cpan}. Moreover, the \glspl{air} of ring constellations are compared to those of \gls{cscg} modulation. Simulations show that 32 rings, 16 \gls{sic}-stages, and Gaussian message passing on the factor graph of the \gls{cpan} surrogate model achieve the \gls{jdd} rates of \gls{cscg} modulation. The computational complexity scales in proportion to the number of \gls{sic}-stages, where one stage has complexity similar to separate detection and decoding.
\end{abstract}

\begin{IEEEkeywords}
Belief propagation, capacity, optical fiber communication, phase noise, successive interference cancellation.
\end{IEEEkeywords}

\glsresetall

\input{introduction}

\input{preliminaries}
\input{SIC_Gaussian}

\input{SIC_Ring}
\input{conclusions}
\section*{Acknowledgment}
\noindent The authors wish to thank Daniel Plabst for inspiring discussions. The authors acknowledge the financial support by the Federal Ministry of Education and Research of Germany in the program “Souverän. Digital. Vernetzt.”. Joint project 6G-life, project identification number: 16KISK002.
\input{appendix}
\bibliographystyle{IEEEtran}
\bibliography{library}
\end{document}

%% file: glossary.tex
\newacronym{acf}{ACF}{autocorrelation function}
\newacronym{air}{AIR}{achievable information rate}
\newacronym{app}{APP}{a posteriori probability}
\newacronym{ase}{ASE}{amplified spontaneous emission}
\newacronym{ask}{ASK}{amplitude-shift keying}
\newacronym{awgn}{AWGN}{additive white Gaussian noise}

\newacronym{b2b}{B2B}{back-to-back}
\newacronym{baa}{BAA}{Blahut-Arimoto Algorithm}
\newacronym{bpcu}{bpcu}{bits per channel use}

\newacronym{cd}{CD}{chromatic dispersion}
\newacronym{cdf}{cdf}{cumulative density function}
\newacronym{coi}{COI}{channel of interest}
\newacronym{cpan}{CPAN}{correlated phase and additive noise}
\newacronym{cscg}{CSCG}{circularly symmetric complex Gaussian}
\newacronym{csi}{CSI}{channel state information}
\newacronym{cwt}{CWT}{continouus-wave tone}

\newacronym{dbp}{DBP}{digital backpropagation}
\newacronym{dsp}{DSP}{digital signal processing}
\newacronym{dtft}{DTFT}{discrete time Fourier transform}

\newacronym{dd}{DD}{direct detection}
\newacronym{dmd}{DMD}{differential mode delay}
\newacronym{dmgd}{DMGD}{differential mode group delay}
\newacronym{dgd}{DGD}{differential group delay}
\newacronym{dt}{DT}{decision tree}

\newacronym{evd}{EVD}{Eigenvalue decomposition}

\newacronym{fmf}{FMF}{few-mode fiber}
\newacronym{fwhm}{FWHM}{full-width half-maximum}
\newacronym{fwm}{FWM}{four-wave mixing}

\newacronym{ghq}{GHQ}{Gauss-Hermite-Quadrature}
\newacronym{gmi}{GMI}{generalized mutual information}
\newacronym{gmp}{GMP}{Gaussian message passing}

\newacronym{idra}{IDRA}{ideal distributed Raman amplification}
\newacronym{iid}{i.i.d.}{independent and identically distributed}
\newacronym{im}{IM}{intensity modulation}
\newacronym{isi}{ISI}{inter-symbol interference}
\newacronym{iss}{ISS}{inter-symbol sample}

\newacronym{jd}{JD}{joint detection}
\newacronym{jdd}{JDD}{joint detection and decoding}
\newacronym{jdabaa}{DABAA}{dynamic-assignment Blahut-Arimoto Algorithm}

\newacronym{kld}{KLD}{Kullback-Leibler-Divergence}
\newacronym{kk}{KK}{Kramers-Kronig}

\newacronym{lstm}{LSTM}{long short-term memory}
\newacronym{lp}{LP}{linearly polarized}

\newacronym{mcf}{MCF}{multi-core fiber}
\newacronym{md}{MD}{modal dispersion}
\newacronym{mgdm}{MGDM}{mode-group-division multiplexing}
\newacronym{mdg}{MDG}{mode-dependent gain}
\newacronym{mdl}{MDL}{mode-dependent loss}
\newacronym{mdm}{MDM}{mode-division multiplexing}
\newacronym{me}{ME}{Manakov equation}
\newacronym{mi}{MI}{mutual information}
\newacronym{mir}{MIR}{mutual information rate}
\newacronym{mimo}{MIMO}{multiple-in multiple-out}
\newacronym{miso}{MISO}{multiple-in single-out}
\newacronym{ml}{ML}{machine learning}
\newacronym{mmf}{MMF}{multi-mode fiber}
\newacronym{mpc}{MPC}{minimum-phase criterion}
\newacronym{mzm}{MZM}{Mach-Zehnder modulator}

\newacronym{nlse}{NLSE}{nonlinear Schrödinger equation}
\newacronym{nn}{NN}{neural network}

\newacronym{pam}{PAM}{pulse-amplitude modulation}
\newacronym{pdf}{pdf}{probability density function}
\newacronym{pmf}{pmf}{probability mass function}
\newacronym{psd}{PSD}{power spectral density}
\newacronym{psk}{PSK}{phase-shift keying}

\newacronym{qam}{QAM}{quadrature-amplitude modulation}

\newacronym{roadm}{ROADM}{reconfigurable optical add-drop multiplexer}
\newacronym{rp}{RP}{regular perturbation}
\newacronym{rrc}{RRC}{root-raised cosine}
\newacronym{rv}{RV}{random variable}

\newacronym{sd}{SD}{seperate detection}
\newacronym{sdd}{SDD}{separate detection and decoding}
\newacronym{sdm}{SDM}{space-division multiplexing}
\newacronym{sic}{SIC}{successive interference cancellation}
\newacronym{simo}{SIMO}{single-in multiple-out}
\newacronym{sinr}{SINR}{signal to interference and noise ratio}
\newacronym{siso}{SISO}{single-in single-out}
\newacronym{sld}{SLD}{square-law detector}
\newacronym{snr}{SNR}{signal-to-noise ratio}
\newacronym{smf}{SMF}{single-mode fiber}
\newacronym{spa}{SPA}{sum-product algorithm}
\newacronym{ssb}{SSB}{single side-band}
\newacronym{ssbi}{SSBI}{signal-signal beat interference}
\newacronym{ssfm}{SSFM}{split-step Fourier method}
\newacronym{ssmf}{SSMF}{standard single-mode fiber}
\newacronym{svd}{SVD}{singular value decomposition}
\newacronym{svm}{SVM}{state vector machine}

\newacronym{te}{TE}{transverse electric}
\newacronym{tm}{TM}{transverse magnetic}

\newacronym{urr}{URR}{unidistant Rayleigh ring}

\newacronym{wdm}{WDM}{wavelength-division multiplexing}
\newacronym{wlg}{w.l.g.}{without loss of generality}

\newacronym{xpm}{XPM}{cross-phase modulation}

%% file: introduction.tex
\section{Introduction}
Estimating the capacity of optical fiber is difficult because of the interactions of attenuation, dispersion, and Kerr non-linearity \cite{Essiambre:10:Capacity}. A standard approach computes \glspl{air} by simulating transmission and having receivers process their signals via surrogate models. The closer the surrogate and actual models, as measured by informational divergence, the higher the \glspl{air}. Past work has increased \glspl{air} by increasing the surrogate model complexity \cite{Essiambre:10:Capacity,Marsella:15:AIR, Secondini:19:Wiener,Gomez:20:CPAN}.

Two useful surrogate models are a memoryless \gls{awgn} channel whose covariance and pseudo-covariance may depend on the channel input amplitude \cite[Sec.~X.C]{Essiambre:10:Capacity} and an \gls{awgn} channel with correlated phase noise and large memory \cite{Mecozzi:12:Pseudolinear,Dar:13:NLIN,Dar:14:Bounds}; see also \cite{Dar:16:Collision,Secondini:19:Wiener,Secondini:22:Sequence,Gomez:20:CPAN,Gomez:21:CPAN,Gomez:22:CPAN}. A memoryless model suggests simple receiver algorithms with \gls{app} processing. The models with memory improve the \gls{air}, but it is unclear how to build receivers. In particular, the receivers in \cite{Secondini:19:Wiener,Secondini:22:Sequence,Gomez:20:CPAN,Gomez:21:CPAN,Gomez:22:CPAN,Dauwels:08:Particle} use particle filters to compute \gls{jdd} rates, but direct implementation of \gls{jdd} is usually too complex for practical systems. This work aims to bridge the gap between memoryless models and models with memory by proposing practical receiver algorithms.

Two classic methods to approach \gls{jdd} performance combine \gls{sdd} with either turbo processing \cite{Douillard:95:iterative} or \gls{sic}. The former approach was applied to Wiener phase noise channels \cite{Dauwels:04:Wiener,Colavolpe:05:Iterative,Shayovitz:16:Wiener} and fiber-optic systems \cite{Alfredsson:19:Iterative}. This method has the disadvantage of requiring dedicated code design to achieve the \gls{jdd} rates, i.e., one should match the code and detector extrinsic information transfer (EXIT) functions \cite{ten2004design}. This reduces flexibility and makes comparing detectors difficult because one must design and analyze different codes. Moreover, turbo processing exchanges \glspl{app} (soft information) between the detector and decoder, which requires additional storage and delay, and pilot insertion may be needed to commence successful algorithm convergence \cite{Colavolpe:05:Iterative}.
We thus focus on \gls{sic} that permits using coded modulation designed for memoryless channels.

This paper is organized as follows. \secref{sec:preliminaries} specifies notation, the system model, and the \gls{cpan} surrogate model of \cite{Gomez:20:CPAN}. The section also reviews the \gls{spa} on factor graphs, \gls{gmi}, and \gls{sic}. \secref{sec:gaussian} and \ref{sec:ring} propose \gls{sic} receivers for \gls{cscg} modulation and ring constellations, respectively. The receivers use belief propagation with \gls{gmp}, i.e., the messages are second-order statistics. For a growing number of \gls{sic}-stages, the receiver achieves and even surpasses the \gls{air} for \gls{jdd} predicted in \cite{Gomez:20:CPAN}. \secref{sec:conclusion} concludes the paper and suggests further work on implementations.

%% file: preliminaries.tex
\section{Preliminaries}
\label{sec:preliminaries}

\subsection{Notation}
\label{subsec:notation}
Random variables are written in uppercase, such as $X$, and their realizations in lowercase, such as $x$. Random vectors are written in bold, such as $\vek{X}$, and their realizations as $\vek{x}$. The \gls{pdf} of $\vek{X}$ is $p_{\vek{X}}(\cdot)$, or simply $p(\cdot)$ if the \gls{pdf} argument, or the context, makes clear to which random vector is being referred to. 
Expectation with respect to $p(\cdot)$ is denoted $\mathrm{E}_p[\vek{X}]$, or simply $\mathrm{E}[\vek{X}]$ if $p(\cdot)$ is the density of $\vek{X}$. The entropy of a discrete-alphabet $\vek{X}$ is $H(\vek{X})$; the differential entropy of a continuous-alphabet $\vek{X}$ is $h(\vek{X})$; the \gls{mi} of $\vek{X}$ and $\vek{Y}$ is $I(\vek{X};\vek{Y})$; the informational divergence of the densities $p(\cdot)$ and $q(\cdot)$, where $p(\cdot)$ is the density of $\vek{X}$, is $D(p(\vek{X})\|q(\vek{Y}))$. Conditioning on  $\vek{Z}$ is written as $H(\vek{X}|\vek{Z})$, $h(\vek{X}|\vek{Z})$, $I(\vek{X};\vek{Y}|\vek{Z})$, and $D(p(\vek{X}|\vek{Z})\|q(\vek{Y}|\vek{Z})|r(\vek{Z}))$ if $\vek{Z}$ has the density $r(\cdot)$. We write $a(x)\propto b(x)$ if $a(x)=c\cdot b(x)$ for some constant $c$.

A Gaussian $X$ with mean $\mu$ and variance $\sigma^2$ has \gls{pdf}
\begin{equation}
    \normpdf{x}{\mu}{\sigma^2} = \frac{1}{\sqrt{2\pi\sigma^2}}\exp\left(-\frac{1}{2}\frac{(x-\mu)^2}{\sigma^2}\right).
\end{equation}
Similarly, a complex Gaussian $X$ with mean $\mu$, variance $\sigma^2=\mathrm{E}\left[|X-\mu|^2\right]$ and pseudo-variance $p^2 = \mathrm{E}\left[(X-\mu)^2\right]$ has \gls{pdf}
\begin{equation}
    \begin{aligned}    
        &\cnormpdf{x}{\mu}{\sigma^2}{p^2} = \frac{1}{\pi\sqrt{\sigma^2\left(\sigma^2-\frac{|p|^4}{\sigma^2}\right)}}\\
        &\exp\left(-\frac{1}{2}[(x-\mu)^*,(x-\mu)]\begin{bmatrix}
            \sigma^2&p^2\\\left(p^2\right)^*&\sigma^2
        \end{bmatrix}^{-1}\begin{bmatrix}
            (x-\mu)\\(x-\mu)^*
        \end{bmatrix}\right)
    \end{aligned}
\end{equation}
where $^*$ denotes complex conjugation. A \gls{cscg} variable has $p^2=0$ and therefore \gls{pdf}
\begin{equation}
    \cscgnormpdf{x}{\mu}{\sigma^2} = \frac{1}{\pi\sigma^2}\exp\left(-\frac{|x-\mu|^2}{\sigma^2}\right).
\end{equation}
The function $\anglebr{x}$ returns the angle of a complex number $x$.
The function 
\begin{equation}
    \modpi{x} = \left((x+\pi) \mod 2\pi\right)-\pi
\end{equation}
maps real numbers to the interval $[-\pi,\pi)$.

\subsection{System Model}
\label{subsec:system_model}

We use a standard model \cite{Essiambre:10:Capacity} for optical networks with co-propagating \gls{wdm} signals. We study single-polarization transmission where the continuous-time, complex-valued, baseband signal for $\nsym$ symbols and $2\nchannels$ interfering \gls{wdm} channels is
\begin{equation}
    x(0,t) = \sum_{i=1}^\nsym x_ig(t-i\tsym)+\sum_{\substack{k=-\nchannels\\k\neq 0}}^\nchannels \sum_{i=1}^\nsym b^{(k)}_ig(t-i\tsym)\e^{\imag \omega_k t}.
    \label{eq:wdm-model}
\end{equation}
The $x_i$ and $b_i^{(k)}$ are the transmit symbols of the \gls{coi} and $k$-th interfering channel, respectively. They are realizations of independent and identically distributed (i.i.d.) zero-mean random variables with alphabet $\mathcal{X}$ and variance $\sigma_x^2$. We use $\mathrm{sinc}$ pulses $g(t) \propto \sin(\pi t/\tsym)/(\pi t/\tsym)$ that are normalized so the per-channel average transmit power is $\ptx = \sigma_x^2$. The symbol rate is $B_\mathrm{ch}=1/\tsym$ and the central frequency of the $k$-th \gls{wdm} channel is $\omega_k/2\pi$. The channel spacing is $B_\mathrm{sp}$ Hz, so $\omega_k=2\pi B_\mathrm{sp}k$.

Signal propagation is governed by the \gls{nlse} \cite{Gomez:20:CPAN}
\begin{equation}
    \frac{\partial x(z,t)}{\partial z} = -\imag \frac{\beta_2}{2}\frac{\partial^2 x(z,t)}{\partial t^2}+\imag\gamma |x(z,t)|^2x(z,t)+n(z,t)
\end{equation}
where  $\beta_2$ is the dispersion coefficient,  $\gamma$ is the nonlinearity coefficient, and $n(z,t)$ is noise dominated by \gls{ase}. Attenuation is removed by \gls{idra}. The \gls{ase} noise spectral density is $N_\mathrm{ASE}=\alpha L hf \eta$, where $\alpha=\alpha_\mathrm{dB}/(10\log_{10} e)$ is the fiber loss coefficient, $L$ is the fiber length, $hf$ is the photon energy at optical frequency $f$, and $\eta$ is the phonon occupancy factor; see~\cite[p.~678]{Essiambre:10:Capacity}.

Co-propagating \gls{wdm} signals interfere if $\gamma\ne0$, e.g., through \gls{xpm} and \gls{fwm}. Each receiver accesses its channel via a bandpass filter with bandwidth $B_\mathrm{ch}$. It then performs sampling, single-channel \gls{dbp}, $\mathrm{sinc}$ filtering, downsampling to the symbol rate, and mean phase rotation compensation \cite{Gomez:20:CPAN} to obtain the sequence $\{y_i\}$. 

\subsection{CPAN-Model}
\label{subsec:cpan}
Surrogate models based on \gls{rp} \cite{Mecozzi:12:RP} simplify computation and analysis. We use the \gls{cpan} model from \cite{Gomez:20:CPAN} that has a phase noise channel
\begin{equation}
\label{equ:cpan_channel}
    Y_i = X_i\e^{\imag\Theta_i}+N_i
\end{equation} 
where the transmit symbols $\{X_i\}$ are \gls{iid}. The additive noise process $\{N_i\}$ is white and \gls{cscg} with  $p(n_i)=\cscgnormpdf{n_i}{0}{\sigma_n^2}$,
and the phase noise process $\{\Theta_i\}$ is a Markov chain with unit memory:
\begin{equation}
    \Theta_i = \incmean \Theta_{i-1} + \incstd \Delta_i
\end{equation}
where $\{\Delta_i\}$ has i.i.d. real-valued, zero-mean, unit-variance, Gaussian $\Delta_i$. We refer to \cite[Equ. (50)]{Gomez:20:CPAN}, \cite[Equ. (56)]{Gomez:20:CPAN} and \eqref{equ:incmeanstd} on how to choose $\incmean$ and $\incstd$. The $\Theta_i$ are zero-mean Gaussian with variance $\thetavar$ for all $i$. The additive and phase noise are independent of the transmit string
\begin{equation}
    \vek{X} = [X_1,X_2,\ldots,X_\nsym].
\end{equation}

Note that the surrogate models are simpler than in \cite{Gomez:20:CPAN}: the surrogate phase noise memory is $1$ rather than $3$, and the surrogate additive noise is white rather than filtered. These modifications simplify the detector and cause a small rate loss. For instance, by comparing \figref{fig:results_gaussian_b} below with \cite[Fig. 5]{Gomez:20:CPAN}, the rate loss is approximately \SI{0.2}{\bpcu}.

\subsection{Factor Graphs and the Sum-Product Algorithm}
\label{sec:spa}

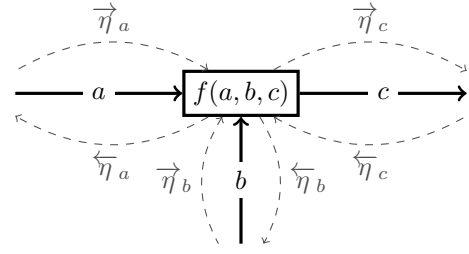
\begin{figure}
    \centering
    \begin{tikzpicture}
        \node[draw, very thick] (f) at (0,0) {$f(a,b,c)$};
        \draw[<-, very thick] (f)--node[midway, fill=white] {$a$} ++(-3cm,0);
        \draw[<-, very thick] (f)--node[midway, fill=white] {$b$} ++(0,-2cm);
        \draw[->, very thick] (f)--node[midway, fill=white] {$c$} ++(3cm,0);
        \draw[msgarrow] (f) edge[bend right, <-] node[above] {$\etaF_{a}$} ++(-3cm,0.3cm);
        \draw[msgarrow] (f) edge[bend left, ->] node[below] {$\etaB_{a}$} ++(-3cm,-0.3cm);

        \draw[msgarrow] (f) edge[bend left, ->] node[above] {$\etaF_{c}$} ++(3cm,0.3cm);
        \draw[msgarrow] (f) edge[bend right, <-] node[below] {$\etaB_{c}$} ++(3cm,-0.3cm);

        \draw[msgarrow] (f) edge[bend right, <-] node[left] {$\etaF_{b}$} ++(-0.3cm,-2cm);
        \draw[msgarrow] (f) edge[bend left, ->] node[right] {$\etaB_{b}$} ++(0.3cm,-2cm);
    \end{tikzpicture}
    \caption{A node $f(a,b,c)$ and its edges $a,b,c$ in a factor graph.}
    \label{fig:fg_example}
\end{figure}

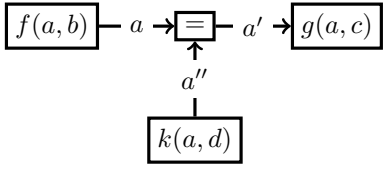
\begin{figure}
    \centering
    \begin{tikzpicture}
        \node[draw, very thick] (equ) at (0,0) {$=$};
        \node[draw, left = 1cm of equ, very thick] (f) {$f(a,b)$};
        \node[draw, right = 1cm of equ, very thick] (g) {$g(a,c)$};
        \node[draw, below = 1cm of equ, very thick] (k) {$k(a,d)$};
        \draw[->, very thick] (f)--node[midway,fill=white]{$a$} (equ);
        \draw[->, very thick] (equ)--node[midway,fill=white]{$a'$} (g);
        \draw[->, very thick] (k)--node[midway,fill=white] {$a''$} (equ);
    \end{tikzpicture}
    \caption{Usage of an equality node.}
    \label{fig:fg_equ}
\end{figure}

A product of probabilities representing a joint probability can be visualized by a factor graph, where nodes represent functions and edges represent variables\cite{Kschischang:01:SPA,Loeliger:04:SPA}; see \figref{fig:fg_example}. Two nodes are connected with an edge if a variable appears in both functions. Equality constraints represent variables that appear in more than two functions. For example, \figref{fig:fg_equ} shows a variable $a$ that appears in the functions $f,g,k$. The edges $a,a',a''$ take the same value. This is ensured by the equality node with the local function $\delta(a-a')\delta(a-a'')$.

We use the \gls{spa} to calculate marginal distributions. A factor graph is usually undirected, but we use directed edges to specify the direction of the \gls{spa} messages. A message $\etaF_a$ in the direction of edge $a$ has a right arrow, and a message $\etaB_a$ in the opposite direction has a left arrow. The dashed edges are not part of the factor graph; we draw them to help visualize the messages, see \figref{fig:fg_example}.

When a function node receives a message from all its edges except one, it computes and passes a message over the remaining edge. For example, once node $f$ in Fig.~\ref{fig:fg_example} has received $\etaF_a$ and $\etaF_b$, it computes $\etaF_c$ and passes it over edge $c$. The calculation is performed according to (see \cite{Kschischang:01:SPA})
\begin{equation}
    \etaF_c(z) = \int_\mathcal{A}\int_\mathcal{B}\etaF_a(x)\etaF_b(y)f(x,y,z) \, \mathrm{d}x\,\mathrm{d}y
    \label{eq:SPA-function-node-compuation}
\end{equation}
for $z \in \mathcal C$, where $\mathcal{A},\mathcal{B},\mathcal{C}$ are the alphabets of $a,b,c$, respectively. Node $f$ similarly computes $\etaB_a$ and $\etaB_b$ once the messages $(\etaF_b$, $\etaB_c)$ and $(\etaF_a$, $\etaB_c)$ arrive, respectively. The \gls{spa} messages for belief propagation are probability density functions, which are usually too complex for implementation. We will approximate densities by Gaussians, i.e., we use \gls{gmp}. For example, for real-valued Gaussians, one passes the mean $\muF_a$ and variance $\sigmaF^2_a$ of $\etaF_a$, and likewise for $\etaB_a$.

As a final step,  marginal distributions are calculated by multiplying the oppositely directed messages of each edge. For example, the marginalization for the variable $a$ in \figref{fig:fg_example} can be calculated using the product $\etaF_a(\cdot)\cdot\etaB_a(\cdot)$.

\subsection{Generalized Mutual Information}
We calculate \glspl{air} via surrogate models; see \cite[Ex. 5.22]{Gallager68}. The receiver chooses a non-negative decoding function $q(\vek{x},\vek{y})$ and, given $\vek{y}$, selects (one of) the $\vek{x}$ that maximizes this function. An \gls{air} is the normalized \gls{gmi} (see \cite{Kaplan:93:Mismatched,Scarlett:20:Mismatched})
\begin{equation}
\label{eq:gmi}
    \frac{1}{\nsym} I_\mathrm{GMI}
    = \frac{1}{\nsym} \max_{\mathfrak s>0} \mathrm{E}\left[\log\frac{q(\vek{X},\vek{Y})^{\mathfrak s}}{\mathrm{E}_{p_{\vek{X}}}\left[ q(\vek{X},\vek{Y})^{\mathfrak s}\right]}\right].
\end{equation}
To understand this expression, one may interpret the argument of the logarithm in \eqref{eq:gmi} as a ratio $q(\vek{y}|\vek{x})/q(\vek{y})$ of probability densities. For example, setting $q(\vek{x},\vek{y})^{\mathfrak s}=p(\vek{y}|\vek{x})$ makes the argument of the logarithm in \eqref{eq:gmi} become $p(\vek{y}|\vek{x})/p(\vek{y})$ so that $I_\mathrm{GMI}=I(\vek{X};\vek{Y})$. The function $q(\vek{x},\vek{y})$ may thus be interpreted as an unnormalized surrogate model $q(\vek{y}|\vek{x})$.

\subsection{Successive Interference Cancellation}
\label{sec:sic}
\Gls{sic} and multistage encoding can bridge the gap between \glspl{air} for models without and with memory. For example, three classes of channels with \emph{block} memory are as follows.
\begin{itemize}[leftmargin=*]
\item Higher-order modulation: consider a $2^m$-ary constellation such as 8-ASK (amplitude shift keying) with $m=3$. One may encode $m$ bitstreams with binary encoders of different rates and map each $m$-tuple of bits (one bit from each encoder output) to a $2^m$-ary symbol. The receiver may apply \gls{sic} to decode, i.e., \gls{sic} stage $s$, $s=1,\dots,m$, decodes bit level $s$ of each symbol. One may interpret the overall bit channel as having memory within blocks of $m$ bits (the $2^m$-ary symbols) and memoryless across blocks.
\item Space-time coding: consider $M$ antennas with $2^m$-ary symbols on each antenna. One may encode $M$ symbol streams, one for each antenna, and then apply \gls{sic} with $M$ stages. Alternatively, each $M$-tuple of $2^m$-ary symbols may be viewed as a $2^{Mm}$-ary symbol. Thus, one may encode and decode using \gls{sic} for $Mm$ bitstreams.
\item Polar codes: consider $k$ data bits that are carefully interlaced with $2^m-k$ zeros to give a vector of $2^m$ bits. Polar coding multiplies this vector with a $2^m\times2^m$ matrix, namely the $m$-fold Kronecker product of a $2\times2$ binary matrix. Decoding is usually performed using \gls{sic} over $k$ stages, where the decoding order is chosen to give a low error probability.
\end{itemize}
In all cases, \gls{sic} permits approaching \gls{jdd} rates using signaling for memoryless channels, i.e., \gls{sic} simplifies encoding and decoding. Historically, \gls{sic} decoding for higher-order modulation was proposed in~\cite{Imai:IT77}. The chain rule of \gls{mi} was used to choose the per-stage code rates in~\cite{Wachsmann:99:Multilevel}; see \eqref{equ:jdd_a_b}-\eqref{equ:jdd_b} below. Similar ideas appeared for space-time coding  in~\cite{Foschini-BLTJ96}. Polar codes with \gls{sic} decoding were developed in \cite{stolte-02,Arikan-IT09}.

Instead of block memory, we are interested in channels with \emph{sliding-window} memory, e.g., \gls{isi} channels. \Gls{sic} decoding for linear \gls{isi} channels was proposed in~\cite{Guess-ISIT00,Pfister:01:ISI,Soriaga:07:ISI}; the applications were copper wire communication and magnetic recording. \Gls{sic} decoding for long-haul and short-reach fiber was proposed in \cite[Sec.~XII]{Essiambre:10:Capacity} and~\cite{Prinz:23:SIC}, respectively; see also~\cite{Plabst:24:SIC,Jaeger:24:ecoc}.

We describe an example of \gls{sic} with $\nstages=2$ stages. Observe that the channel may have block, sliding-window, or any other type of memory. For example, the channel need not be causal, i.e., the channel response may have temporal precursors and postcursors for each transmit symbol. The transmitted vector $\vek{x}$ of even dimension $\nsym$ is divided into $\nstages=2$ vectors $\vek{a}$ and $\vek{b}$ of dimension $\nsym/2$ as follows:
\begin{equation}
\label{equ:splitting_x}
    \vek{x} = [a_1,b_1,a_2,b_2,\ldots a_{\nsym/2},b_{\nsym/2}].
\end{equation}
The chain rule of \gls{mi} gives
\begin{align}    
    I(\vek{X};\vek{Y}) = I(\vek{A};\vek{Y})+I(\vek{B};\vek{Y}|\vek{A})
    \label{equ:jdd_a_b}
\end{align}
and with independent $X_i$ for all $i$, we have
\begin{align}    
    \label{equ:jdd_a}I(\vek{A};\vek{Y}) &= \sum_{i=1}^{n/2} h(A_i)-h(A_i|\vek{Y},A_1,\ldots,A_{i-1})\\
    \label{equ:jdd_b}I(\vek{B};\vek{Y}|\vek{A})&=\sum_{i=1}^{n/2}h(B_i)-h(B_i|\vek{Y},\vek{A},B_1,\ldots,B_{i-1}).
\end{align}

\begin{figure}
    \centering
    \begin{tikzpicture}
        \node[] (y) at (0,0) {$\vek{y}$};
        \node[below = 1.5cm of y] (det_center) {};
        \node[draw, left = 1cm of det_center] (detectora) {Detector 1};
        \node[draw, below = 1.5 of detectora] (decodera) {Decoder 1};
        \node[draw, right = 1cm of det_center] (detectorb) {Detector 2};
        \node[draw, below = 1.5 of detectorb] (decoderb) {Decoder 2};
        \node[below = 0.5cm of decodera] (outputa) {$\vek{a}$};
        \node[below = 0.5cm of decoderb] (outputb) {$\vek{b}$};
        
        \draw[->] (y) -- (detectora);
        \draw[->] (y) -- (detectorb);
        \draw[->] (detectora) -- node[midway, fill={white}] {\small $p(a_i|\vek{y})$} (decodera);
        \draw[->] (detectorb) -- node[midway, fill={white}] {\small $p(b_i|\vek{y},\vek{a})$} (decoderb);
        \draw[->] (decodera) -- node[midway, fill={white}] {\small $\vek{a}$} (detectorb);
        \draw[->] (decodera)--(outputa);
        \draw[->] (decoderb)--(outputb);
    \end{tikzpicture}
    \caption{\gls{sic} with two stages.}
    \label{fig:sic}
\end{figure}
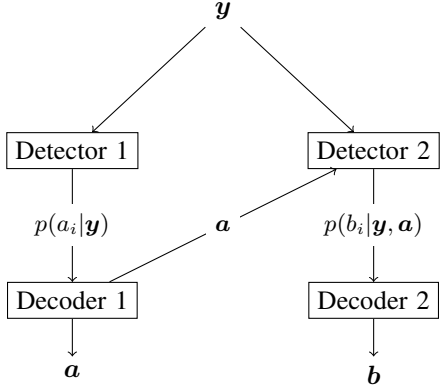

Given a received vector $\vek{y}$, the \gls{sic} detector works in $\nstages=2$ stages, see \figref{fig:sic}:
\begin{enumerate}
    \item Compute the symbol-wise \glspl{app} $p(a_i|\vek{y})$ for all $i$. Decoder 1 uses these to make a decision $\hat{\vek{a}}$ on $\vek{a}$.
    \item Compute the symbol-wise \glspl{app} $p(b_i|\vek{y},\hat{\vek{a}})$ for all $i$. Decoder 2 uses these to make a decision $\hat{\vek{b}}$ on $\vek{b}$.
\end{enumerate}
If the transmission rate for the first stage is less than its \gls{air}, then there exist codes with error probability arbitrarily close to zero; we thus assume $\hat{\vek{a}}=\vek{a}$. A similar statement can be made for the second-stage estimate $\hat{\vek{b}}$ of $\vek{b}$.

Observe that the decoder receives \glspl{app} from the detector without information on the inter-symbol dependencies. Hence, $p(a_i|\vek{y})$ is commonly treated as being independent of the $p(a_k|\vek{y})$ with $k\neq i$. For decoding stage $s=1$, a \gls{gmi} under this independence assumption is (note the \gls{mi} subscript)
\begin{align}
\label{equ:sic_1_mi}
    I_1(\vek{A};\vek{Y})
    := \sum_{i=1}^{\nsym/2} I(A_i;\vek{Y})
    = \sum_{i=1}^{\nsym/2} h(A_i) - h(A_i|\vek{Y}).
\end{align}
To see this, insert $\vek{x}=\vek{a}$, $p(\vek{a})=\prod_i p(a_i)$, and $q(\vek{a},\vek{y})^{\mathfrak s}=\prod_i p(\vek{y}|a_i)$ into \eqref{eq:gmi}. Moreover, comparing \eqref{equ:sic_1_mi} and \eqref{equ:jdd_a}, and using
$h(A_i|\vek{Y})\ge h(A_i|\vek{Y},A_1,\dots,A_{i-1})$, we have $I_1(\vek{A};\vek{Y}) \le I(\vek{A};\vek{Y})$.

Similarly, a \gls{gmi} for the decoding stage $s=2$ is
\begin{equation}
\label{equ:sic_2_mi}
    I_2(\vek{B};\vek{Y}|\vek{A}) := \sum_{i=1}^{\nsym/2} h(B_i)-h(B_i|\vek{Y},\vek{A}).
\end{equation}
Comparing \eqref{equ:sic_2_mi} and \eqref{equ:jdd_b}, we have $I_2(\vek{B};\vek{Y}|\vek{A}) \le I(\vek{B};\vek{Y}|\vek{A})$ because $h(B_i|\vek{Y},\vek{A})\ge h(B_i|\vek{Y},\vek{A},B_1,\dots,B_{i-1})$. Combining \eqref{equ:jdd_a_b} and \eqref{equ:sic_1_mi}--\eqref{equ:sic_2_mi}, an \gls{air} for \gls{sic} is
\begin{align}
    \frac{1}{\nsym}I_\mathrm{sic}(\vek{X};\vek{Y}) &:= \frac{1}{\nsym}\Big(I_1(\vek{A};\vek{Y})+ I_2(\vek{B};\vek{Y}|\vek{A})\Big) \nonumber \\
    &\le \frac{1}{\nsym}I(\vek{X};\vek{Y}).
    \label{equ:MI_sic_inequality}
\end{align}

For $\nstages>2$, assume $\nstages$ divides $\nsym$ and split $\vek{x}$ into $\nstages$ vectors
\begin{equation}
  \vek{x}^{(s)} :=[{x^{(s)}_1},{x^{(s)}_2},\ldots,{x^{(s)}_{\nsym/\nstages}}], \quad s=1,\dots,S .
\end{equation}
We interlace these vectors as
\begin{equation} \label{eq:canonic-arrangement}
    \vek{x} = [ {x^{(1)}_1,\ldots , x^{(\nstages)}_1},{x^{(1)}_2,\ldots x_2^{(\nstages)}},\ldots ,{x^{(1)}_{\nsym/\nstages},\ldots , x^{(\nstages)}_{\nsym/\nstages}} ]
\end{equation}
to reduce the channel memory; observe that each stage decodes only every $\nstages$-th transmitted symbol. This implies that in stage $s$, $1<s\le\nstages$, the $s-1$ temporal precursors to any symbol are decoded. Other symbol arrangements may also be used.

%% file: SIC_Gaussian.tex
\section{SIC for Gaussian Inputs}
\label{sec:gaussian}

Consider \gls{cscg} inputs with $p(x_i) = \cscgnormpdf{x_i}{0}{\sigma_x^2}$. Such continuous modulation is impractical but useful for performance analysis and system design. For example, the results extend to discrete modulation formats, as shown in Sec.~\ref{sec:ring} for discrete amplitudes and in \cite{Jaeger:24:ecoc} for fully discrete modulation.

\subsection{Surrogate \Gls{app} Based on the CPAN Model}
We write $p_X(\cdot)$ and $q_X(\cdot)$ for true and surrogate \glspl{pdf}, respectively. As described in \secref{subsec:notation}, we remove the subscript if the random variable is clear from the context.

The detector wishes to compute $p(a_i|\vek{y})$ and $p(b_i|\vek{y},\vek{a})$. However, the true pdfs are unavailable, so we use a detector for the surrogate model \eqref{equ:cpan_channel} with density
\begin{align}
    q(\vek{x},\vek{y},\vek{\theta}) &= p(\vek{x}) q(\vek{\theta})q(\vek{y}|\vek{x},\vek{\theta}) \nonumber\\
    &=\prod_{i=1}^\nsym p(x_i)q(\theta_i|\theta_{i-1})q(y_i|x_i,\theta_i)
    \label{equ:mismatched_prob}
\end{align}
where we applied $q(\vek{y}|\vek{x},\vek{\theta})=\prod_{i=1}^nq(y_i|x_i,\theta_i)$ and where
\begin{align}
    p(x_i) &= \cscgnormpdf{x_i}{0}{\sigma_x^2}\\
    q(\theta_i|\theta_{i-1}) &= \normpdf{\theta_i}{\incmean\theta_{i-1}}{\incvar}\\
    q(\theta_1) &= \normpdf{\theta_1}{0}{\thetavar}\\ 
    q(y_i|x_i,\theta_i) &= \cscgnormpdf{y_i}{x_i\e^{\imag \theta_i}}{\sigma_n^2}.
\end{align}
Consider $S=2$ stages and recall from \eqref{equ:splitting_x} that $\vek{x}$ consists of $\vek{a}$ and $\vek{b}$. We will approximate $p(a_i|\vek{y})$ and $p(b_i|\vek{y},\vek{a})$ by the respective
\begin{align}
    q(a_i|\vek{y})&=\frac{1}{c_1}\int_{\mathbb{R}^n}\int_{\mathcal{A}^{\backslash \{i\}}} q(\vek{x},\vek{y},\vek{\theta})\,\mathrm{d}\vek{x}\,\mathrm{d}\vek{\theta}\\
    q(b_i|\vek{y},\vek{a})&=\frac{1}{c_2}\int_{\mathbb{R}^n}\int_{\mathcal{B}_{\vek{a}}^{\backslash \{i\}}} q(\vek{x},\vek{y},\vek{\theta})\, \mathrm{d}\vek{x}\,\mathrm{d}\vek{\theta}
\end{align}
where $c_1$ and $c_2$ are normalization factors and
\begin{align}
    \mathcal{A}^{\backslash\{i\}} &= \left\{\vek{x}\in\mathbb{C}^\nsym:x_{2i-1}=a_i\right\}\\
    \mathcal{B}_{\vek{a}}^{\backslash\{i\}} &= \left\{\vek{x}\in\mathbb{C}^\nsym:x_{2i}=b_i,[x_1,x_3,\ldots,x_{n-1}]=\vek{a}\right\} .
\end{align}

\subsection{Efficient Computation of the Marginal Distributions}
We marginalize $q(\vek{x},\vek{y},\vek{\theta})$ in both \gls{sic}-stages by using the \gls{spa}; see \secref{sec:spa}.

\subsubsection{First Stage Detection}
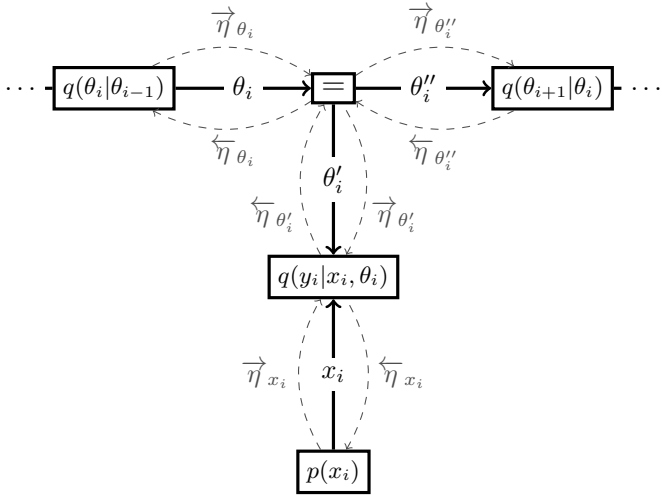
\begin{figure}
    \centering
    \begin{tikzpicture}
    \node[draw, shape = rectangle, very thick] (ptheta1) at (0,0) {\small$q(\theta_i|\theta_{i-1})$};
    \node[draw, shape = rectangle, right = 1.8cm of ptheta1, very thick] (equ1) {\large$=$};
    \node[draw, shape = rectangle, right = 1.8cm of equ1, very thick] (ptheta2){\small$q(\theta_{i+1}|\theta_i)$};
    
    \node[draw, shape = rectangle, below = 2cm of equ1, very thick] (qy1) {\small $q(y_i|x_i,\theta_i)$};
    \node[draw, shape = rectangle, below = 2cm of qy1, very thick] (px1) {\small$p(x_i)$};

    \draw[->, very thick] (ptheta1)--(equ1) node[midway,fill=white] {$\theta_i$};
    \draw[->, very thick] (equ1)--(ptheta2) node[midway,fill=white] {$\theta_i''$};
    
    \draw[->, very thick] (equ1)--(qy1) node[midway,fill=white] {$\theta_i'$};
    \draw[->, very thick] (px1)--(qy1) node[midway,fill=white] {$x_i$};

    \draw[] (ptheta1) edge[msgarrow, bend left, ->] node[above] {$\etaF_{\theta_i}$} (equ1);
    \draw[] (equ1) edge[msgarrow, bend left, ->] node[above] {$\etaF_{\theta_i''}$} (ptheta2);
    
    \draw[] (equ1) edge[msgarrow, bend right, <-] node[below] {$\etaB_{\theta_i''}$} (ptheta2);
    \draw[] (ptheta1) edge[msgarrow, bend right, <-] node[below] {$\etaB_{\theta_i}$} (equ1);
    
    \node[left = 0.1cm of ptheta1] (leftdots) {$\ldots$};
    \node[right = 0.1cm of ptheta2] (rightdots) {$\ldots$};
    \draw[very thick] (ptheta2)--(rightdots);
    \draw[very thick] (leftdots)--(ptheta1);
    
    \draw[] (px1) edge[msgarrow, bend left, ->] node[left] {$\etaF_{x_i}$} (qy1);
    
    \draw[] (px1) edge[msgarrow, bend right, <-] node[right] {$\etaB_{x_i}$} (qy1);
    
    \draw[] (qy1) edge[msgarrow, bend left, ->] node[left, pos=1/4] {$\etaB_{\theta_i'}$} (equ1);
    
    \draw[] (qy1) edge[msgarrow, bend right, <-] node[right, pos=1/4] {$\etaF_{\theta_i'}$} (equ1);
    
    \end{tikzpicture}
    \caption{Branch of the non-decoded stage. The equality constraint implies $\theta_i=\theta_i'=\theta_i''$. However, the messages of these variables are different in general.}
    \label{fig:first_stage}
\end{figure}

\figref{fig:first_stage} depicts the graph of the first \gls{sic}-stage based on \eqref{equ:mismatched_prob}. The equality constraint appears because $\theta_i$ is a variable in three functions; see \secref{sec:spa}.

\smallskip\noindent
\textbf{Upward Path:}
We have $\etaF_{x_i}(\cdot)=p_X(\cdot)$. The $X_i$ are circularly symmetric, i.e., we have $p_{X}(x) = p_X(x\e^{\imag\theta})$ for all $x$ and $\theta$, which implies
\begin{align}
\label{equ:thetaB_p}
    \etaB_{\theta'_i}(\theta) &= \normB{\theta_i'}\int_\mathbb{C} p(x)q(y_i|x,\theta)\,\mathrm{d}x \nonumber \\
    &= \normB{\theta_i'}\int_\mathbb{C} p(x')q(y_i|x',0)\,\mathrm{d}x' \;= \const
\end{align}
where $x'=x\e^{\imag\theta}$. By $\normB{\theta_i'}$, and likewise for other messages, we denote a constant that normalizes to a valid \gls{pdf}.

The message $\etaB_{\theta'_i}$ represents a surrogate density of $y_i$ conditioned on $\theta_i$. As this is constant in $\theta_i$, knowledge of the instantaneous phase noise does not alter the probability of the output. This is intuitive, provided the transmitted phase $\angle x_i$ is uniformly distributed and added to $\theta_i$ according to \eqref{equ:cpan_channel}.

\smallskip\noindent
\textbf{Rightward Path:}
Using $\etaF_{\theta_1}(\cdot)=q_{\theta_1}(\cdot)=q_\theta(\cdot )$, we obtain
\begin{align}    
    \etaF_{\theta_2}(\theta) &=\normF{\theta_2}\int_\mathbb{R} \etaF_{\theta_1}(\phi)\etaB_{\theta_1'}(\phi) q_{\theta_2|\theta_1}(\theta|\phi)\,\mathrm{d}\phi \nonumber \\
    &= \int_\mathbb{R} q_{\theta}(\phi)q_{\theta_2|\theta_1}(\theta|\phi)\,\mathrm{d}\phi= q_{\theta}(\theta)
\end{align}
where the second step follows because $\etaB_{\theta'_1}(\theta)$ is constant in $\theta$ and therefore cancels the normalization constant $\normF{\theta_2}$. We recursively obtain $\etaF_{\theta_i}(\cdot ) = q_{\theta}(\cdot )$ for all $i$.

\smallskip\noindent
\textbf{Leftward Path:} We similarly have
\begin{equation}
    \begin{aligned}    
        \etaB_{\theta''_{\nsym-1}}(\theta) &=\normB{\theta''_{\nsym-1}}\int_\mathbb{R}\etaB_{\theta_\nsym'}(\phi)q_{\theta_n|\theta_{n-1}}(\phi|\theta)\,\mathrm{d}\phi
        = \const
    \end{aligned}
\end{equation}
and recursively $\etaB_{\theta''_i}(\theta)$ is constant in $\theta$ for all $i$.

\smallskip\noindent
\textbf{Downward Path:} We have 
\begin{equation}
\label{equ:cscg_etaF_thetai_first_stage}
    \etaF_{\theta_i'}(\theta) = \normF{\theta_i'}\etaF_{\theta_i}(\theta)\etaB_{\theta_i''}(\theta) = q(\theta)
\end{equation}
and
\begin{align} 
    \etaB_{x_i}(x) &= \normB{x_i}\int_\mathbb{R} \etaF_{\theta'_i}(\theta)q(y_i|x,\theta)\,\mathrm{d}\theta \nonumber\\    &=\normB{x_i}\int_\mathbb{R}\normpdf{\theta}{0}{\thetavar}\cscgnormpdf{y_i}{x\e^{\imag\theta}}{\sigma_n^2}\,\mathrm{d}\theta.
    \label{equ:muB_x}
\end{align}
The surrogate \gls{app} $q(x_i|\vek{y})$ may now be calculated using
\begin{align}
    f_i(x) = \frac{1}{c_{f_i}}\etaF_{x_i}(x)\etaB_{x_i}(x) \label{equ:fi}
\end{align}
where $c_{f_i}$ normalizes $f_i$ to a valid \gls{pdf}.

The integral \eqref{equ:muB_x} seems to have no closed-form expression. \Gls{gmp} approximates $f_i(\cdot)$ by a complex Gaussian with mean $\mu_{f_i}=\mathrm{E}_{f_i}[X]$, variance $\sigma_{f_i}^2=\mathrm{E}_{f_i}\left[|X-\mu_{f_i}|^2\right]$ and pseudo-variance $p_{f_i}^2=\mathrm{E}_{f_i}\left[(X-\mu_{f_i})^2\right]$. The quality of the simulation results below justifies the approximation. As derived in \appref{app:derivation_f}, we thus have
\begin{equation}
\label{equ:q_x_first_stages}
    q(x_i|\vek{y}) = \cnormpdf{x_i}{\mu_{f_i}}{\sigma_{f_i}^2}{p_{f_i}^2}
\end{equation}
with
\begin{align}
\label{equ:mu_x_first_stages}    \mu_{f_i} &= y_i\frac{\sigma_x^2}{\sigma_y^2}\exp\left(-\frac{\thetavar}{2}\right)\\
\label{equ:sigma_x_first_stages}    \sigma_{f_i}^2 &= \frac{\sigma_x^2}{\sigma_y^2}\left(\sigma_n^2+|y_i|^2\frac{\sigma_x^2}{\sigma_y^2}\right)-\left|\mu_{f_i}\right|^2\\
\label{equ:p_x_first_stages}  p_{f_i}^2 &= y_i^2\frac{\sigma_x^4}{\sigma_y^4}\exp\left(-2\thetavar\right)-\mu_{f_i}^2
\end{align}
where $\sigma_y^2=\sigma_x^2+\sigma_n^2$. At this point, we are interested only in $q(x_i|\vek{y})$ for odd $i$, as these are the symbols detected in the first \gls{sic}-stage. Observe that $q(x_i|y_i)= q(x_i|\vek{y})$, so the first stage uses a memoryless detector conditioned on $y_i$ only.

\subsubsection{Second Stage Detection}
\begin{figure}
    \centering
    \begin{tikzpicture}
        \node[draw, shape = rectangle, very thick] (ptheta1) at (0,0) {\small$q(\theta_i|\theta_{i-1})$};
        \node[draw, shape = rectangle, right = 1.8cm of ptheta1, very thick] (equ1) {\large$=$};
        \node[draw, shape = rectangle, right = 1.8cm of equ1, very thick] (ptheta2){\small$q(\theta_{i+1}|\theta_i)$};
        
        \node[draw, shape = rectangle, below = 2cm of equ1, very thick] (qy1) {\small $p(x_i)q(y_i|x_i,\theta_i)$};

        \draw[->, very thick] (ptheta1)--(equ1) node[midway,fill=white] {$\theta_i$};
        \draw[->, very thick] (equ1)--(ptheta2) node[midway,fill=white] {$\theta_i''$};
        
        \draw[->, very thick] (equ1)--(qy1) node[midway,fill=white] {$\theta_i'$};

        \draw[msgarrow] (ptheta1) edge[bend left, ->] node[above] {$\etaF_{\theta_i}$} (equ1);
        \draw[msgarrow] (equ1) edge[bend left, ->] node[above] {$\etaF_{\theta_i''}$} (ptheta2);
        
        \draw[msgarrow] (equ1) edge[bend right, <-] node[below] {$\etaB_{\theta_i''}$} (ptheta2);
        \draw[msgarrow] (ptheta1) edge[bend right, <-] node[below] {$\etaB_{\theta_i}$} (equ1);
        
        \node[left = 0.1cm of ptheta1] (leftdots) {$\ldots$};
        \node[right = 0.1cm of ptheta2] (rightdots) {$\ldots$};
        \draw[very thick] (ptheta2)--(rightdots);
        \draw[very thick] (leftdots)--(ptheta1);
        
        \draw[msgarrow] (qy1) edge[bend left, ->] node[left, pos=1/4] {$\etaB_{\theta_i'}$} (equ1);
        
        \draw[msgarrow] (qy1) edge[bend right, <-] node[right, pos=1/4] {$\etaF_{\theta_i'}$} (equ1);
    \end{tikzpicture}
    \caption{Branch of the decoded stage.}
    \label{fig:second_stage}
\end{figure}
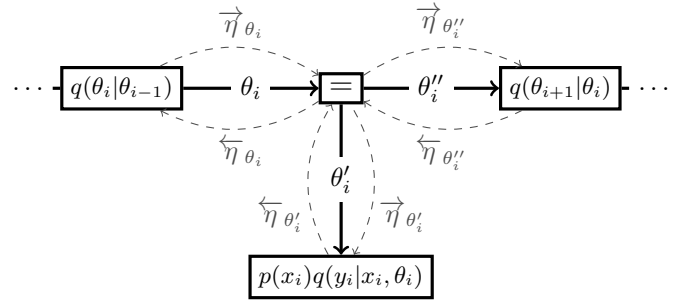

Suppose the symbols in $\vek{x}$ with an odd index $i$, namely those described by $\vek{a}$, have been correctly decoded. Hence, branches of the form in \figref{fig:first_stage} and branches of the form in \figref{fig:second_stage} alternate. The former correspond to the elements in $\vek{b}$ or those with even index of $\vek{x}$, respectively, and the latter to those in $\vek{a}$ or odd index of $\vek{x}$.

\smallskip\noindent
\textbf{Upward Path:} For odd $i$, the message passed over $\theta_i'$ is
\begin{align}  
    \etaB_{\theta_i'}(\theta) &= \normB{\theta_i'}p(x_i)q(y_i|x_i,\theta) \nonumber \\
    &= \normB{\theta_i'}\frac{p(x_i)}{\pi \sigma_n^2}\exp\left(-\frac{\left|y_i-x_i\e^{\imag\theta}\right|^2}{\sigma_n^2}\right) \nonumber \\
    &\propto \exp\left(\frac{2|y_i||x_i|}{\sigma_n^2}\cos\Big(\theta-\left(\anglenobr{y_i}-\anglenobr{x_i}\right)\Big)\right).
    \label{equ:periodicity_argument}
\end{align}
One can replace $\anglenobr{y_i}-\anglenobr{x_i}$ with $\modpi{\anglenobr{y_i}-\anglenobr{x_i}}$ since this adds an integer multiple of $2\pi$ to the cosine argument. From \eqref{equ:cpan_channel}, for small variances of the \gls{awgn} and phase noise, the resulting cosine argument is small and using $\cos(x)\approx 1-x^2/2$ gives
\begin{align}
\label{equ:etaBthetaip}
    \etaB_{\theta_i'}(\theta) &\approx \normpdf{\theta}{\muB_{\theta_i'}}{\sigmaB^2_{\theta_i'}}\\
    \muB_{\theta_i'} &= \modpi{\anglenobr{y_i}-\anglenobr{x_i}}\\
    \sigmaB^2_{\theta_i'} &=\frac{\sigma_n^2}{2|y_i||x_i|}.
\end{align}
For even $i$, the message $\etaB_{\theta_i'}(\theta)$ is constant in $\theta$; see \eqref{equ:thetaB_p}. 

\smallskip\noindent
\textbf{Rightward Path:}
We show that all messages in the rightward path are approximately Gaussian, that is
\begin{align}
\label{equ:etaFthetai}    \etaF_{\theta_i}(\theta)&\approx\normpdf{\theta}{\muF_{\theta_i}}{\sigmaF^2_{\theta_i}}\\
\label{equ:etaFthetaipp}    \etaF_{\theta_i''}(\theta)&\approx\normpdf{\theta}{\muF_{\theta_i''}}{\sigmaF^2_{\theta_i''}}.
\end{align}
If $\etaF_{\theta_i}$ is Gaussian, then $\etaF_{\theta_i''}$ is Gaussian, since it is either a product of Gaussians or a product of a Gaussian and a constant \cite{Bromiley:14:Gaussians}. Explicitly, the parameters of $\etaF_{\theta_i''}$ depend on $i$ as follows.
\begin{itemize}
    \item If $i$ is odd, then $x_i$ was already decoded in the first stage and is a branch of the form shown in \figref{fig:second_stage}. Hence $\muF_{\theta_i''}$ is a product of Gaussians and \cite{Bromiley:14:Gaussians}
    \begin{align}
        \muF_{\theta_i''} &= \frac{\muF_{\theta_i}\sigmaB^2_{\theta_i'}+\muB_{\theta_i'}\sigmaF^2_{\theta_i}}{\sigmaF^2_{\theta_i}+
        \sigmaB^2_{\theta_i'}}\\
        \sigmaF^2_{\theta_i''} &= \frac{\sigmaF^2_{\theta_i}\sigmaB^2_{\theta_i'}}{\sigmaF^2_{\theta_i}+\sigmaB^2_{\theta_i'}}.
    \end{align}
    \item If $i$ is even, then $x_i$ was not decoded in the first stage and is a branch of the form shown in \figref{fig:first_stage}. Hence, $\etaF_{\theta_i''}$ is a product of a Gaussian and a constant, and therefore $\muF_{\theta_i''}=\muF_{\theta_i}$ and $\sigmaF^2_{\theta_i''}=\sigmaF^2_{\theta_i}$.
\end{itemize}
 If $\etaF_{\theta_{i-1}''}$ is Gaussian, then $\etaF_{\theta_i}$ is the marginalization over the product of a Gaussian and a conditional Gaussian. We have
\begin{align}
    \int_\mathbb{R}\normpdf{\theta_{i-1}}{\mu}{\sigma^2}\normpdf{\theta_i}{\incmean\theta_{i-1}}{\incvar}\,\mathrm{d}\theta_{i-1} \nonumber \\
    =\normpdf{\theta_i}{\incmean\mu}{\incmean^2\sigma^2+\incvar}
    \label{equ:markovupdateintegral}
\end{align}
and therefore 
\begin{align}
    \etaF_{\theta_i}(\theta) &= \int_\mathbb{R} \etaF_{\theta_{i-1}''}(\phi)q_{\theta_i|\theta_{i-1}}(\theta|\phi)\,\mathrm{d}\phi \nonumber \\
    &\approx\normpdf{\theta}{\incmean\muF_{\theta''_{i-1}}}{\incmean^2\sigmaF^2_{\theta_{i-1}''}+\incvar}.
\end{align}
Using $\etaF_{\theta_1}(\theta)=q_{\theta_1}(\theta)=\normpdf{\theta}{0}{\thetavar}$ for any stage, we arrive at \eqref{equ:etaFthetai}--\eqref{equ:etaFthetaipp} by induction.

\smallskip\noindent
\textbf{Leftward Path:}
Let $i'$ be the largest index of the symbols decoded in earlier stages. For example, for $S=2$ we have $i'=\nsym-1$. All branches to the right of the $i'$-th branch have the form shown in \figref{fig:first_stage} and therefore $\etaB_{\theta_{i'}''}(\theta)$ is constant in $\theta$. We find that (note the subscripts)
\begin{equation}
    \etaB_{\theta_{i'}}(\theta) = \etaB_{\theta_{i'}'}(\theta)
\end{equation}
is approximately Gaussian; see \eqref{equ:etaBthetaip}. If $\etaB_{\theta_{i+1}}(\theta)$ is Gaussian in $\theta$, then we update
\begin{align}
    \etaB_{\theta_i''}(\theta) &= \int_\mathbb{R} \etaB_{\theta_{i+1}}(\phi)q_{\theta_{i+1}|\theta_i}(\phi|\theta)\,\mathrm{d}\phi \nonumber \\
    &\approx \normpdf{\theta}{\frac{\muB_{\theta_{i+1}}}{\incmean}}{\frac{\sigmaB^2_{\theta_{i+1}}+\incvar}{\incmean^2}}.
    \label{equ:etaBthetipp}
\end{align}
Similar to the rightward path, for $i\leq i'$, we have
\begin{equation}
    \etaB_{\theta_i}(\theta) \approx \normpdf{\theta}{\muB_{\theta_i}}{\sigmaB^2_{\theta_i}}
\end{equation}
where the update rule depends on the index $i$.
\begin{itemize}
    \item If $i$ is odd, then 
    \begin{align}
        \muB_{\theta_i} &= \frac{\muB_{\theta_i''}\sigmaB^2_{\theta_i'}+\muB_{\theta_i'}\sigmaB^2_{\theta_i''}}{\sigmaB^2_{\theta_i'}+\sigmaB^2_{\theta_i''}}\\
        \sigmaB^2_{\theta_i} &= \frac{\sigmaB^2_{\theta_i'}\sigmaB^2_{\theta_i''}}{\sigmaB^2_{\theta_i'}+\sigmaB^2_{\theta_i''}}.
    \end{align}
    \item If $i$ is even, then $\muB_{\theta_i}=\muB_{\theta_i''}$ and $\sigmaB^2_{\theta_i}=\sigmaB^2_{\theta_i''}$.
\end{itemize}

\smallskip\noindent
\textbf{Downward Path:}
As both $\etaF_{\theta_i}(\theta)$ and $\etaB_{\theta_i''}(\theta)$ are Gaussian in $\theta$, their product is also Gaussian. That is, we have
\begin{align}  
        \etaF_{\theta_i'}(\theta) &=\normF{\theta_i'}\etaF_{\theta_i}(\theta)\etaB_{\theta_i''}(\theta) \nonumber \\
        &= \normpdf{\theta}{\muF_{\theta_i'}}{\sigmaF^2_{\theta_i'}}
\end{align}
with
\begin{align}
\label{equ:muFthetap}\muF_{\theta_i'} &= \frac{\muF_{\theta_i}\sigmaB^2_{\theta_i''}+\muB_{\theta_i''}\sigmaF^2_{\theta_i}}{\sigmaF^2_{\theta_i}+\sigmaB^2_{\theta_i''}} \\
\label{equ:sigmaFtheap}\sigmaF^2_{\theta_i'} &= \frac{\sigmaF^2_{\theta_i}\sigmaB^2_{\theta_i''}}{\sigmaF^2_{\theta_i}+\sigmaB^2_{\theta_i''}}.
\end{align}
Similar to the first stage, for even $i$, we approximate $q(x_i|\vek{y},\vek{a})$ by a complex Gaussian. Simulations show that using \glspl{cscg} suffices, and the mean and variance are (see \appref{app:derivation_f})
\begin{align}
    \label{equ:mu_f} \mu_{f_i} &= y_i\frac{\sigma_x^2}{\sigma_y^2}\exp\left(-\frac{1}{2}\frac{\muF_{\theta'_i}^2-(\muF_{\theta_i'}-\imag\sigmaF^2_{\theta_i'})^2}{\sigmaF^2_{\theta_i'}}\right)\\
    \label{equ:sigma_f} \sigma_{f_i}^2 &= \frac{\sigma_x^2}{\sigma_y^2}\left(\sigma_n^2+|y_i|^2\frac{\sigma_x^2}{\sigma_y^2}\right)-\left|\mu_{f_i}\right|^2.
\end{align}

\subsection{Extension to $\nstages$ \gls{sic}-Stages}
An extension to $\nstages$ stages is straightforward. The first stage operates as described by \eqref{equ:q_x_first_stages}--\eqref{equ:p_x_first_stages}. For stage $s>1$, all $x_i$ corresponding to stages $s'<s$ are assumed to be known, and the derivation is similar to that of stage $s=2$.

The following means and variances should be calculated beforehand for appropriate indices $i$:
\begin{align}
    \muB_{\theta_i'} = \modpi{\anglenobr{y_i}-\anglenobr{x_i}}, \quad
    \sigmaB^2_{\theta_i'} = \frac{\sigma_n^2}{2|y_i||x_i|}.
\end{align}
For \gls{gmp}, we collect the mean and variance in one vector
\begin{equation}
    \etaFvec_{\theta_i} = \left[\muF_{\theta_i},\sigmaF^2_{\theta_i}\right]
\end{equation}
and likewise for other messages. We also define the function
\begin{equation}
    g(\vek{\eta}_1,\vek{\eta}_2) = \left[
        \frac{\mu_1\sigma_2^2+\mu_2\sigma_1^2}{\sigma_1^2+\sigma_2^2},
        \frac{\sigma_1^2\sigma_2^2}{\sigma_1^2+\sigma_2^2}
    \right]
\end{equation}
which describes the mean and variance of the product of Gaussians with parameters $\vek{\eta}_1$ and $\vek{\eta}_2$.

Algorithm \ref{alg:gaussian} shows the computations for stage $s$. The set $\mathcal{I}_s$ has the indices of symbols decoded in earlier stages, e.g., for $\nstages=2$ we have $\mathcal{I}_1=\emptyset$ and $\mathcal{I}_2=\{1,3,\ldots,\nsym-1\}$. We further have $\etaFvec_{\theta_1} = [0,\thetavar]$. For $i'=\nsym-\nstages+(s-1)$, which is the index of the last symbol in $\vek{x}$ decoded prior to stage $s>1$, we obtain
\begin{equation}
    \etaBvec_{\theta_{i'}} = \left[\modpi{\anglenobr{y_{i'}}-\anglenobr{x_{i'}}},\frac{\sigma_n^2}{2|y_{i'}||x_{i'}|}\right]
\end{equation}
and (see \eqref{equ:etaBthetipp})
\begin{equation}
    \etaBvec_{\theta_{i'-1}''} = \left[\frac{\modpi{\anglenobr{y_{i'}}-\anglenobr{x_{i'}}}}{\incmean},\frac{\frac{\sigma_n^2}{2|y_{i'}||x_{i'}|}+\incvar}{\incmean^2}\right].
\end{equation}
Let $\vek{x}^{(s)}$ be the symbols decoded in stage $s$, i.e., for two stages $\vek{a}=\vek{x}^{(1)}$ and $\vek{b}=\vek{x}^{(2)}$. For $i\in \{s,s+\nstages,s+2\nstages,\ldots\}$, we have
\begin{equation}
\label{equ:a-posteriori}
    q(x_i|\vek{y},\vek{x}^{(1)},\ldots,\vek{x}^{(s-1)})
    = \cscgnormpdf{x_i}{\mu_{f_i}}{\sigma_{f_i}^2}
\end{equation}
where $\mu_{f_i}$ and $\sigma_{f_i}^2$ can be calculated from the output of Algorithm \ref{alg:gaussian} and \eqref{equ:mu_f}-\eqref{equ:sigma_f}. We remark that the calculations for the $\etaBvec_{\theta_i'}$, $\etaFvec_{\theta_i'}$, $\mu_{f_i}$, and $\sigma_{f_i}^2$ can be parallelized.

\begin{algorithm}[t]
\caption{\gls{sic}-Stage-Detector for \gls{cscg} Inputs}\label{alg:gaussian}
\textbf{Input:} $\vek{y}$, $\vek{x}^{(1)},\ldots,\vek{x}^{(s-1)}$, $\mathcal{I}_s$, $s$, $\nstages$, $\nsym$, $i'$,$\etaFvec_{\theta_1}$, $\etaBvec_{\theta_{i'-1}''}$, $\etaBvec_{\theta_i'}$ for $i\in\mathcal{I}_s$\\
\textbf{Output:} $\etaFvec_{\theta_i'}$
\begin{algorithmic}
\For{$i\gets 1$ to $\nsym-1$}
\algorithmiccomment{Rightward Path}
\If{$i\in\mathcal{I}_s$}
    \State $\etaFvec_{\theta_i''}\gets g(\etaFvec_{\theta_i},\etaBvec_{\theta_i'})$
\Else
    \State $\etaFvec_{\theta_i''}\gets\etaFvec_{\theta_i}$    
\EndIf
\State $\etaFvec_{\theta_{i+1}}\gets\left[\incmean\muF_{\theta_i''},\incmean^2\sigmaF^2_{\theta_i''}+\incvar\right]$
\EndFor
\For{$i\gets i'-1$ to $2$}
\algorithmiccomment{Leftward Path}
\If{$i\in\mathcal{I}_s$}
    \State $\etaBvec_{\theta_i}\gets g(\etaBvec_{\theta_i''},\etaBvec_{\theta_i'})$    
\Else
    \State $\etaBvec_{\theta_i}\gets \etaBvec_{\theta_i''}$
\EndIf
\State $\etaBvec_{\theta_{i-1''}}\gets\left[\frac{\muB_{\theta_i}}{\incmean},\frac{\sigmaB^2_{\theta_i}+\incvar}{\incmean^2}\right]$
\EndFor
\For{$l\gets 0$ to $\lfloor{\nsym/\nstages}\rfloor-1$}
\algorithmiccomment{Downward Path}
\State $i\gets s+l\nstages$
\State $\etaFvec_{\theta_i'}\gets g(\etaFvec_{\theta_i},\etaBvec_{\theta_i''})$
\EndFor
\end{algorithmic}
\end{algorithm}

\subsection{Lower Bounds on Mutual Information}
\label{subsec:lower_bound_mi}
We describe how to lower-bound the \gls{sic} \glspl{air} in \eqref{equ:sic_1_mi}-\eqref{equ:MI_sic_inequality}. Consider the steps
\begin{align} 
    h_q(A_i|\vek{Y})&:=-\int_{\mathbb{C}^\nsym} p(\vek{y})\int_\mathbb{C} p(a_i|\vek{y})\log q(a_i|\vek{y})\,\mathrm{d}a_i\,\mathrm{d}\vek{y} \nonumber \\
    &=h(A_i|\vek{Y})+D\left(p(A_i|\vek{Y})||q(A_i|\vek{Y})|p(\vek{Y})\right) \nonumber \\
    &\geq h(A_i|\vek{Y}).
    \label{equ:h_q_a}
\end{align}
A lower bound on $I_1(\vek{A};\vek{Y})$ in \eqref{equ:sic_1_mi} is thus
\begin{equation}
    I_{1,q}(\vek{A};\vek{Y}) := \sum_{i=1}^{\nsym/2} h(A_i)-h_q(A_i|\vek{Y})\leq I_1(\vek{A};\vek{Y}).
\end{equation}
Similarly, a lower bound on $I_2(\vek{B};\vek{Y}|\vek{A})$ in \eqref{equ:sic_2_mi}
\begin{equation}
    I_{2,q}(\vek{B};\vek{Y}|\vek{A}) := \sum_{i=1}^{\nsym/2} h(B_i)-h_q(B_i|\vek{Y},\vek{A})\leq I_2(\vek{B};\vek{Y}|\vek{A})
\end{equation}
with 
\begin{align}
    h_q(B_i|\vek{Y},\vek{A}) &:= -\int_{\mathbb{C}^{\nsym/2}}\int_{\mathbb{C}^\nsym} p(\vek{y},\vek{a})\int_\mathbb{C} p(b_i|\vek{y},\vek{a}) \nonumber \\
    & \qquad \quad \cdot\log q(b_i|\vek{y},\vek{a})\,\mathrm{d}b_i\,\mathrm{d}\vek{y}\,\mathrm{d}\vek{a}.
    \label{equ:h_q_b} 
\end{align}

Next, observe that \gls{cscg} inputs with variance $\sigma_x^2$ have
\begin{equation}
    h(A_i)=h(B_i) = \log(\pi\e\sigma_x^2).
\end{equation} We approximate $h_q(A_i|\vek{Y})$ in \eqref{equ:h_q_a} and $h_q(B_i|\vek{Y},\vek{A})$ in \eqref{equ:h_q_b} by simulating with $\nsequence$ sequences $\{\vek{x}_k\}$ and $\{\vek{y}_k\}$ and computing (see~\cite{Arnold:06:simulation})
\begin{align}
    h_q(A_i|\vek{Y})&\approx -\frac{1}{\nsequence}\sum_{k=1}^\nsequence \log q(a_{k,{i}}|\vek{y}_k)\\
    h_q(B_i|\vek{Y},\vek{A}) &\approx -\frac{1}{\nsequence}\sum_{k=1}^\nsequence \log q(b_{k,i}|\vek{y}_k,\vek{a}_k).   
\end{align}

\subsection{Complexity}

The \gls{sic} complexity scales linearly with $\nstages$ because all stages have comparable complexity. Stage $s=1$ coincides with \gls{sdd} for which the complexity scales linearly in $\nsym$.

For stage $s$ with $s>1$, consider Algorithm \ref{alg:gaussian}. The rightward path is a \emph{for} loop with $\nsym-1$ iterations, each calculating two new messages, which gives $2\nsym-2$ messages. The leftward path has at most $\nsym-1$ iterations, hence $2\nsym-2$ messages. This is followed by the downward path, which creates $\nsym/\nstages$ messages. Finally, $\nsym/\nstages$ \glspl{app} are calculated; see \eqref{equ:a-posteriori}. The input of Algorithm \ref{alg:gaussian} requires computing $|\mathcal{I}_s|+2$ messages. Since $\nstages$ divides $\nsym$, we have $|\mathcal{I}_s|=(s-1)\frac{\nsym}{\nstages}$ and obtain
\begin{equation}
    2(2\nsym-2)+2\frac{\nsym}{\nstages}+(\nsym-1)\frac{\nsym}{\nstages}+2 = \left(6-\frac{2}{s}\right)n-2
\end{equation}
messages. Therefore, the complexity scales linearly with $\nsym$.

The initialization and computation of the \glspl{app} from the output can be parallelized, and so can all computations on the downward path. The rightward and leftward paths may run in parallel, but the for-loop within each path cannot be parallelized, as it uses results from the previous iteration.

\subsection{Estimating Parameters}
\label{subsec:est_params}
 The phase noise variance is \cite{Dar:13:NLIN}
\begin{equation}
 \thetavar = \frac{4\gamma^2 L}{\tsym}\sum_{\substack{k=-\nchannels\\k\neq 0}}^\nchannels  \frac{\kurtosis-\sigma_x^4}{|\beta_2\omega_k|}
\end{equation} 
where $\kurtosis$ denotes the kurtosis of the modulation alphabet. The parameters of the increment are \cite{Gomez:20:CPAN}
\begin{equation}
\label{equ:incmeanstd}    \incmean = \frac{r}{\thetavar}\qquad\incstd^2 = \thetavar-\frac{r^2}{\thetavar}
\end{equation}
with 
\begin{equation}
    r = \frac{4\gamma^2 L }{\tsym}\sum_{\substack{k=-\nchannels\\k\neq 0}}^\nchannels  \frac{\kurtosis-\sigma_x^4}{|\beta_2\omega_k|}\max\left(0,1-\frac{\tsym}{|\beta_2\omega_k|L}\right).
\end{equation}
We use $\ntraining$ training sequences $\{\vek{x}_k\}$ and $\{\vek{y}_k \}$ from simulations and estimate the remaining parameters as in \cite{Gomez:20:CPAN,Secondini:19:Wiener}. The variance of the additive noise is
\begin{equation}
    \sigma_n^2 = \arg \max_{\sigma^2} \sum_{k=1}^\ntraining\sum_{i=1}^\nsym \log L\left(\left|y_{k,i}\right|,\left|x_{k,i}\right|;\sigma^2\right)
\end{equation}
with the Rice distribution
\begin{equation}
L(a,b;\sigma^2) = \frac{2a}{\sigma^2}\e^{-\frac{a^2+b^2}{\sigma^2}}I_0\left(\frac{2ab}{\sigma^2}\right)
\end{equation}
where $I_0(\cdot)$ is the modified Bessel function of the first kind of order zero.
Unlike the \gls{cpan}-model in \secref{subsec:cpan}, the phase noise after matched filtering and downsampling has a non-zero mean. We thus compute the estimate \cite{Gomez:20:CPAN}
\begin{equation}
    \hat{\theta} = \anglebr{\frac{1}{\nsym\ntraining}\sum_{k=1}^\ntraining\sum_{i=1}^\nsym y_{k,i}x_{k,i}^*}
\end{equation}
and multiply the output of single-channel \gls{dbp} by $\exp(-\hat{\theta})$.

\subsection{Simulation Results}
\label{subsec:sim_result_Gaussian}

\begin{figure*}[t!]
    \centering
    \begin{tikzpicture}[every text node part/.style={align=center}]
        \node[] (x) at (0,0) {$\vek{x}$};
        \node[draw, right = 0.5cm of x] (ps) {sinc};
        \coordinate[right = 2cm of ps] (channel_center);
        \node[above = 0.5cm of channel_center] (cpan_blocker) {};
        \node[draw, right = 0.8cm of cpan_blocker] (cpan) {CPAN};
        \node[below = 0.5cm of channel_center,shape=circle, inner sep = 0] (add) {$\oplus$};
        \node[draw, right = 0.8cm of add] (ssfm) {SSFM};
        \node[draw, right = 0.6cm of ssfm] (lp) {Lowpass \&\\Down-\\sampling};
        \node[draw, right = 0.5cm of lp] (dbp) {DBP};
        \coordinate[above = 0.5cm of dbp] (dbp_center);
        \node[draw, right = 0.6cm of dbp_center] (mf) {sinc};
        \node[draw, right = 0.5cm of mf] (ds) {Down-\\sampling};
        \node[draw, right = 0.5cm of ds] (mean) {$\cdot\e^{-\hat{\theta}}$};
        \node[draw, right = 0.5cm of mean] (det) {Detector};
        \node[right = 0.5cm of det] (q) {$q(\vek{x}|\vek{y})$};
        \node[below = 0.5cm of add] (interf) {Interfering\\Channels};
        \node[below = 0.5cm of ssfm] (addnoise) {Additive\\Noise};

        \draw[->] (x) -- (ps);
        \draw[->] (lp)--(dbp);
        \draw[->] (dbp.east)-|(mf.south);
        \draw[->] (mf)--(ds);
        \draw[->] (ds)--(mean);
        \draw[->] (mean)--node[midway, above] {$\vek{y}$} (det);
        \draw[->] (det)--(q);
        \draw[->] (cpan.east)-|(mf.north);
        \draw[->] (ssfm.east)--(lp.west);
        \draw[<-] (cpan.west)-|++(-1.9cm,-0.5cm);
        \draw[->] (add)--(ssfm);
        \draw[<-] (add.west)-|++(-0.8cm,0.5cm);
        \node[draw, shape = circle, fill=black, right = 0.5cm of ps, inner sep = 0] (circ) {};
        \draw[] (ps.east)--(circ);
        \draw[] (circ)--++(0.4cm,0.1cm);
        \coordinate[right = .1cm of circ] (arc_center);
        \coordinate[below = .2cm of arc_center] (arc_bottom);
        \draw[] (arc_bottom) arc (-30:30:0.5cm);
        \draw[->] (interf)--(add);
        \draw[->] (addnoise)--(ssfm);

        \draw[dashed, color = matlab1] (-0.4,-1)--(-0.4,1)--(1.9,1)--(1.9,-1)--(-0.4,-1);
        \node[anchor = north, color = matlab1] at (0.75,1) {\small Transmitter};
        \draw[dashed, color = matlab2] (2.7,1.7)--(5.9,1.7)--(5.9,-2.5)--(2.7,-2.5)--(2.7,1.7);
        \node[anchor = north, color = matlab2] at (4.4,1.7) {\small Channel};
        \draw[dashed, color = matlab3] (6,-1.7)--(6,1.7)--(17.7,1.7)--(17.7,-1.7)--(6,-1.7);
        \node[anchor = north, color = matlab3] at (11.85,1.7) {\small Receiver};
    \end{tikzpicture}
    \caption{Signal propagation via the \gls{cpan} model or the \gls{ssfm}. Oversampling accounts for spectral broadening. The receiver applies filtering and downsampling to one sample/symbol before the detector. The \gls{ssfm} block adds noise due to \gls{idra}.}
    \label{fig:simsetup}
\end{figure*}
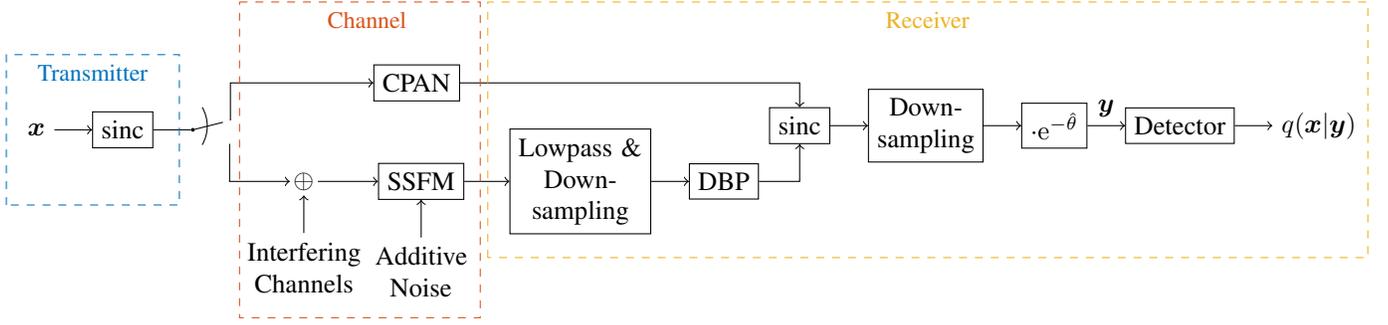

\figref{fig:simsetup} shows the simulation setup. The top path shows a benchmark: the \gls{cpan} model with noise variances that mimic those of the fiber-optic channel. The bottom path shows the fiber-optic channel simulated using the \gls{ssfm}, \gls{idra}, five \gls{wdm} channels, lowpass filtering, and single-channel \gls{dbp}. \tabref{tab:sim_parameters} lists the simulation parameters. We apply $\ntraining=24$ sequences with $8192$ symbols to estimate $\sigma_n^2$ and $\hat{\theta}$ as described in \secref{subsec:est_params}. We then use $\nsequence=120$ sequences with $8192$ symbols each to estimate the \gls{air} as described in \secref{subsec:lower_bound_mi}.

\begin{table}[t!]
    \caption{System Parameters}
    \begin{tabular}{|c|c|c|}
        \hline
        \textbf{Parameter} & \textbf{Symbol} & \textbf{Value}\\
        \hline
        Fiber Length & $L$ & \SI{1000}{\kilo\meter}\\
        Optical Freq. (Wavelength) & $f$ & $\SI{193.414}{\tera\hertz}$ (\SI{1550}{\nano\meter}) \\
        Attenuation & $\alpha_\mathrm{dB}$ & $\SI{0.2}{\dB\per\kilo\meter}$ \\
        Phonon Occupancy Factor & $\eta$ & 1 \\
        Noise Spectral Density & $N_\mathrm{ASE}$ & \SI{5.902e-18 }{\W\per\hertz}\\
        Dispersion Coefficient & $\beta_2$ & \SI{-21.7}{\pico\second\squared\per\kilo\meter}\\
        Nonlinear coefficient & $\gamma$ & \SI{1.27}{\per\watt\per\kilo\meter}\\
        Number of WDM channels & $2C+1$ & 5\\
        Baud Rate, Channel Spacing & $B_\mathrm{ch}$, $B_\mathrm{sp}$ & \SI{50}{\giga\hertz}\\
        \hline
    \end{tabular}
    \label{tab:sim_parameters}
\end{table}

\input{plots/params_ptx}

Consider first the \gls{cpan} channel, i.e., the top path in \figref{fig:simsetup}. \figref{fig:params} plots the variances $\thetavar$, $\incvar$ of the phase noise process, and the variance $\sigma_n^2$ of the \gls{awgn}. The variances increase with $\ptx$ to mimic nonlinear interference. The variance $\sigma_\mathrm{ASE}^2$ of the \gls{ase} is constant at $2.951\cdot 10^{-7}$.

\figref{fig:results_gaussian_a} shows the \glspl{air} in \gls{bpcu} for the following benchmarking scenarios:
\begin{enumerate}[label=\arabic*)] 
\item \label{enum:sdd_awgn} a memoryless \gls{awgn} surrogate model,
\item \label{enum:sdd_pn} a memoryless surrogate model with i.i.d. Gaussian phase noise and independent \gls{awgn},
\item \label{enum:jdd} a \gls{jdd}-receiver based on particle filtering \cite{Secondini:19:Wiener,Gomez:20:CPAN}, and
\item \label{enum:pn} a genie-aided receiver with perfect knowledge of the phase noise, so the \gls{air} is $I(\vek{X};\vek{Y}|\vek{\Theta})$.
\end{enumerate}
Observe that the \glspl{air} increase with the number of \gls{sic} stages and approach the \gls{jdd} rate $I(\vek{X};\vek{Y})$; cf. \eqref{equ:MI_sic_inequality}.
Moreover, we have $I(\vek{X};\vek{Y})\le I(\vek{X};\vek{Y}|\vek{\Theta})$. We remark that the genie-aided \gls{mi} is for a memoryless \gls{awgn} channel where the additive noise variance $\sigma_n^2$ increases with $\ptx$; see \figref{fig:params}.
The solid black curve shows the \gls{awgn} channel capacity with \gls{ase} only, which upper bounds the CPAN channel capacity.
The dashed lines in \figref{fig:air_v_stages} plot the maximum \gls{sic} \glspl{air} over all launch powers $\ptx$. 
\figref{fig:results_gaussian_a} and \figref{fig:air_v_stages} show that \gls{sic} with 2 and 4 stages loses significant \gls{air} compared to \gls{jdd}. One needs at least 8 stages to maintain a rate loss of less than \SI{1}{\%}.

\input{plots/Gaussian_air}

\input{plots/air_v_stages}

Studies of idealized models of dispersion-free nonlinear fiber show that the \glspl{air} grow as ${\frac{1}{2}\log(\mathrm{SNR})+\mathcal{O}(1)}$ where ${\mathrm{SNR}\propto\ptx}$; see \cite{Turitsyn:03:DispFree,Yousefi:11:DispFree,Kramer:18:IT,Häger:22:DispFree}. In contrast, the \gls{cpan} \glspl{air} decrease with $\ptx$ because the additive noise variance $\sigma_n^2$ increases with $\ptx$; see \figref{fig:params}. Thus, both the phase and amplitude of the signal experience distortions that increase with $\ptx$. 

\figref{fig:results_gaussian_b} shows the \glspl{air} for the nonlinear fiber-optic channel with a receiver that uses the \gls{cpan} surrogate model. The solid curve again shows the capacity of the \gls{awgn}-channel distorted by \gls{ase} only, which upper bounds the capacity \cite{Kramer:15:UpperBound,Yousefi:15:UpperBound}. Note that the inequality \eqref{equ:MI_sic_inequality} does not hold for the surrogate \glspl{air}, i.e., the \gls{air} of \gls{sic} might exceed the \gls{air} of \gls{jdd} in \cite{Gomez:20:CPAN}. 
One can increase \glspl{air} by improving the surrogate model, e.g., by including correlations in the additive noise \cite{Secondini:19:Wiener}.
Thus, the \gls{jdd} \glspl{air} are slightly smaller than those in \cite{Gomez:20:CPAN}, mainly because there is no whitening filter.

\figref{fig:results_gaussian_b} and the solid lines in \figref{fig:air_v_stages}  show that 8 \gls{sic}-stages provide \glspl{air} similar to those of \gls{jdd}. Observe that the \gls{sic} \gls{air} slightly exceeds the \gls{jdd} \gls{air} for 16 or more stages. We infer that the true channel is better approximated by the \gls{sic} surrogate channels than the \gls{jdd} surrogate channel. The 64-stage \gls{sic}-receiver gains approximately \SI{0.52}{\bpcu}, or \SI{6.4}{\%}, in rate over the memoryless \gls{awgn} receiver.

%% file: plots/params_ptx.tex
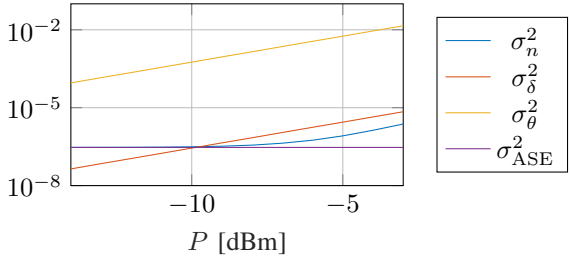
\begin{figure}[t!]
    \centering
        \begin{tikzpicture}    
            \begin{axis}[%
                width=6cm,
                height=4cm,
                at={(0,0)},
                xmin=-14,
                xmax=-3,
                xlabel style={font=\color{white!15!black}},
                xlabel={$\ptx$ [dBm]},
                ymode=log,
                ymin=1e-08,
                ymax=0.1,
                yminorticks=true,
                axis background/.style={fill=white},
                xmajorgrids,
                ymajorgrids,
                legend style={at={(1.1,0.5)},anchor=west,nodes={scale=1, transform shape}},
                legend columns = 1
                ]
                \addplot [color=matlab1]
                  table[]{plots/params_over_ptx-1.tsv};
                  \addlegendentry{$\sigma_n^2$};
                \addplot [color=matlab2]
                  table[]{plots/params_over_ptx-2.tsv};
                  \addlegendentry{$\sigma_\delta^2$};
                \addplot [color=matlab3]
                  table[]{plots/params_over_ptx-3.tsv};
                  \addlegendentry{$\sigma_\theta^2$};
                \addplot [color=matlab4]
                  table[]{plots/params_over_ptx-4.tsv};
                  \addlegendentry{$\sigma_\mathrm{ASE}^2$};
            \end{axis}
        \end{tikzpicture}
    \caption{Parameters of the \gls{cpan} model for the setup described by \tabref{tab:sim_parameters} and \cite[Sec. VIII]{Gomez:20:CPAN}.}
    \label{fig:params}
\end{figure}

%% file: plots/Gaussian_air.tex
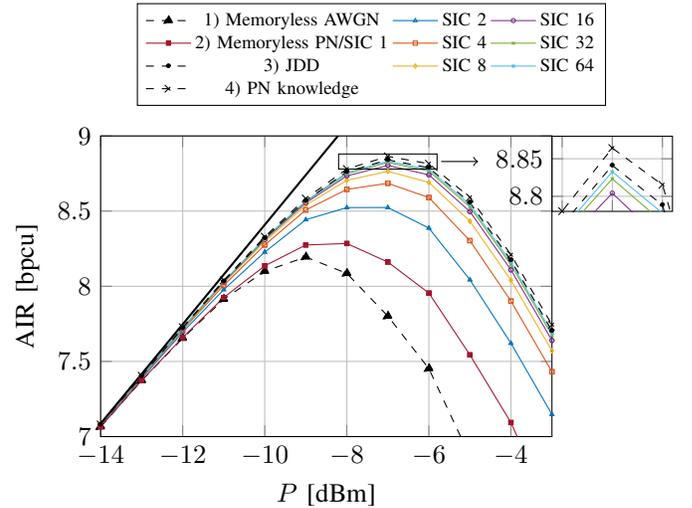
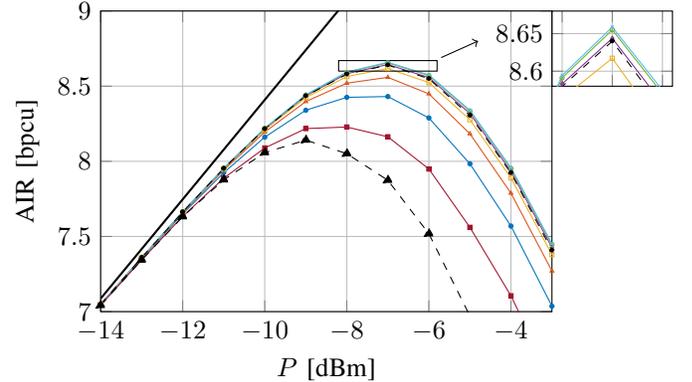
\begin{figure}[t!]
    \centering
    \begin{subfigure}{\linewidth}
        \begin{tikzpicture}    
            \begin{axis}[%
                width=6cm,
                height=4cm,
                at={(0,0)},
                scale only axis,
                xmin=-14,
                xmax=-3,
                ymin=7,
                ymax=9,
                xlabel = {$\ptx$ [dBm]},
                ylabel = {AIR [bpcu]},
                grid = major,
                axis background/.style={fill=white},
                legend style={at={(0.6,1.1)},anchor=south,nodes={scale=0.7, transform shape}},
                legend columns = 3
                ]
                
                \addplot [dashed, color=black, mark = triangle*]
                  table[]{plots/test_data_memoryless_lb-1.tsv};
                 \addlegendentry{\ref{enum:sdd_awgn} Memoryless AWGN};
                \addplot [color=matlab1, draw=none,mark = triangle, mark size = .8pt]
                  table[]{plots/Gaussian_AIR_Test_Data-1.tsv};
                  \addlegendentry{SIC 2};
                \addplot [color=matlab4, draw=none, mark = o, mark size = .8pt]
                  table[]{plots/Gaussian_AIR_Test_Data-4.tsv};
                  \addlegendentry{SIC 16};  
                \addplot [color=matlab7,draw=none, mark = square*, mark size = .8pt]
                  table[]{plots/Gaussian_AIR_Test_Data-9.tsv};
                  \addlegendentry{\ref{enum:sdd_pn} Memoryless PN/\gls{sic} 1 };
                \addplot [color=matlab2, draw=none,mark = square, mark size = .8pt]
                  table[]{plots/Gaussian_AIR_Test_Data-2.tsv};
                  \addlegendentry{SIC 4};
                \addplot [color=matlab5, draw=none, mark = x, mark size = .8pt]
                  table[]{plots/Gaussian_AIR_Test_Data-5.tsv};
                  \addlegendentry{SIC 32};
                \addplot [dashed, color=black, draw=none, mark = *, mark size = .8pt]
                  table[]{plots/Gaussian_AIR_Test_Data-8.tsv};
                  \addlegendentry{\ref{enum:jdd} JDD};                  
                \addplot [color=matlab3, draw=none, mark = diamond, mark size = .8pt]
                  table[]{plots/Gaussian_AIR_Test_Data-3.tsv};
                  \addlegendentry{SIC 8};
                \addplot [color=matlab6, draw=none, mark = asterisk, mark size = .8pt]
                  table[]{plots/Gaussian_AIR_Test_Data-6.tsv};
                  \addlegendentry{SIC 64};  
                \addplot [dashed, color=black, mark = x]
                  table[]{plots/air_test_data_known_phase-1.tsv};
                  \addlegendentry{\ref{enum:pn} PN knowledge};

                \addplot [color=matlab1, forget plot, mark = triangle, mark size = .8pt]
                  table[]{plots/Gaussian_AIR_Test_Data-1.tsv};
                \addplot [color=matlab2, forget plot, mark = square, mark size = .8pt]
                  table[]{plots/Gaussian_AIR_Test_Data-2.tsv};
                \addplot [color=matlab3, forget plot, mark = diamond, mark size = .8pt]
                  table[]{plots/Gaussian_AIR_Test_Data-3.tsv};
                \addplot [color=matlab4, forget plot, mark = o, mark size = .8pt]
                  table[]{plots/Gaussian_AIR_Test_Data-4.tsv};
                \addplot [color=matlab5, forget plot, mark = x, mark size = .8pt]
                  table[]{plots/Gaussian_AIR_Test_Data-5.tsv};
                \addplot [color=matlab6, forget plot, mark = asterisk, mark size = .8pt]
                  table[]{plots/Gaussian_AIR_Test_Data-6.tsv};
                \addplot [color=matlab7, forget plot, mark = square*, mark size = .8pt]
                  table[]{plots/Gaussian_AIR_Test_Data-9.tsv};
                \addplot [dashed, color=black, forget plot, mark = *, mark size = .8pt]
                  table[]{plots/Gaussian_AIR_Test_Data-8.tsv};

                \addplot [color=black, forget plot, style=thick]
                  table[]{plots/air_upper_bound_ase-1.tsv};
                  \draw[] (-8.2,8.78)--(-8.2,8.88)--(-5.8,8.88)--(-5.8,8.78)--(-8.2,8.78);
                  \draw[->] (-5.7,8.83)--(-4.8,8.83);
            \end{axis}
            \begin{axis}[%
                width=1.6cm,
                height=1cm,
                at={(6cm,3cm)},
                scale only axis,
                xmin=-8.2,
                xmax=-5.8,
                ymin=8.78,
                ymax=8.88,
                ytick = {8.8,8.85},
                xticklabel = \empty,
                grid = major,
                axis background/.style={fill=white}
                ]
                \addplot [color=matlab1, forget plot, mark = triangle, mark size = .8pt]
                  table[]{plots/Gaussian_AIR_Test_Data-1.tsv};
                \addplot [color=matlab2, forget plot, mark = square, mark size = .8pt]
                  table[]{plots/Gaussian_AIR_Test_Data-2.tsv};
                \addplot [color=matlab3, forget plot, mark = diamond, mark size = .8pt]
                  table[]{plots/Gaussian_AIR_Test_Data-3.tsv};
                \addplot [color=matlab4, forget plot, mark = o, mark size = .8pt]
                  table[]{plots/Gaussian_AIR_Test_Data-4.tsv};
                \addplot [color=matlab5, forget plot, mark = x, mark size = .8pt]
                  table[]{plots/Gaussian_AIR_Test_Data-5.tsv};
                \addplot [color=matlab6, forget plot, mark = asterisk, mark size = .8pt]
                  table[]{plots/Gaussian_AIR_Test_Data-6.tsv};
                \addplot [color=matlab7, forget plot, mark = square*, mark size = .8pt]
                  table[]{plots/Gaussian_AIR_Test_Data-9.tsv};
                \addplot [dashed, color=black, forget plot, mark = *, mark size = .8pt]
                  table[]{plots/Gaussian_AIR_Test_Data-8.tsv};
                \addplot [dashed, color=black, forget plot, mark = x]
                  table[]{plots/air_test_data_known_phase-1.tsv};
            \end{axis}
        \end{tikzpicture}
        \caption{\gls{cpan} channel.}
        \label{fig:results_gaussian_a}
    \end{subfigure}
    \begin{subfigure}{\linewidth}
        \begin{tikzpicture}
            \begin{axis}[%
                width=6cm,
                height=4cm,
                at={(0,0cm)},
                scale only axis,
                xmin=-14,
                xmax=-3,
                ymin=7,
                ymax=9,
                ylabel = {AIR [bpcu]},
                xlabel = {$\ptx$ [dBm]},
                grid = major,
                axis background/.style={fill=white}
                ]
                    \addplot [color=matlab1, forget plot, mark = *, mark size = .8pt]
                      table[]{plots/Gaussian_AIR-1.tsv};
                    \addplot [color=matlab2, forget plot, mark = triangle, mark size = .8pt]
                      table[]{plots/Gaussian_AIR-2.tsv};
                    \addplot [color=matlab3, forget plot, mark = square, mark size = .8pt]
                      table[]{plots/Gaussian_AIR-3.tsv};
                    \addplot [color=matlab4, forget plot, mark = diamond, mark size = .8pt]
                      table[]{plots/Gaussian_AIR-4.tsv};
                    \addplot [color=matlab5, forget plot, mark = o, mark size = .8pt]
                      table[]{plots/Gaussian_AIR-5.tsv};
                    \addplot [color=matlab6, forget plot, mark = x, mark size = .8pt]
                      table[]{plots/Gaussian_AIR-6.tsv};
                    \addplot [dashed, color=black, forget plot, mark = *, mark size = .8pt]
                      table[]{plots/Gaussian_AIR-9.tsv};
                    \addplot [color=matlab7, forget plot, mark = square*, mark size = .8pt]
                      table[]{plots/Gaussian_AIR-8.tsv};
                    \addplot [color=black, forget plot, style=thick]
                      table[]{plots/air_upper_bound_ase-1.tsv};
                    \addplot [dashed, color=black, forget plot, mark = triangle*]
                      table[]{plots/memoryless_lb-1.tsv};
                      \draw[] (-8.2,8.6)--(-8.2,8.67)--(-5.8,8.67)--(-5.8,8.6)--(-8.2,8.6);
                      \draw[->] (-5.7,8.69)--(-4.8,8.8);
            \end{axis}
            \begin{axis}[%
                width=1.6cm,
                height=1cm,
                at={(6cm,3cm)},
                scale only axis,
                xmin=-8.2,
                xmax=-5.8,
                ymin=8.58,
                ymax=8.68,
                ytick = {8.6,8.65},
                xticklabel = \empty,
                grid = major,
                axis background/.style={fill=white},
                ]
                    \addplot [color=matlab1, forget plot, mark = *, mark size = .8pt]
                      table[]{plots/Gaussian_AIR-1.tsv};
                    \addplot [color=matlab2, forget plot, mark = triangle, mark size = .8pt]
                      table[]{plots/Gaussian_AIR-2.tsv};
                    \addplot [color=matlab3, forget plot, mark = square, mark size = .8pt]
                      table[]{plots/Gaussian_AIR-3.tsv};
                    \addplot [color=matlab4, forget plot, mark = diamond, mark size = .8pt]
                      table[]{plots/Gaussian_AIR-4.tsv};
                    \addplot [color=matlab5, forget plot, mark = o, mark size = .8pt]
                      table[]{plots/Gaussian_AIR-5.tsv};
                    \addplot [color=matlab6, forget plot, mark = x, mark size = .8pt]
                      table[]{plots/Gaussian_AIR-6.tsv};
                    \addplot [dashed, color=black, forget plot, mark = *, mark size = .8pt]
                      table[]{plots/Gaussian_AIR-9.tsv};
                    \addplot [color=matlab7, forget plot, mark = square*, mark size = .8pt]
                      table[]{plots/Gaussian_AIR-8.tsv};
            \end{axis}
        \end{tikzpicture}
        \caption{Nonlinear fiber-optic channel.}
        \label{fig:results_gaussian_b}
    \end{subfigure}       
    \caption{\glspl{air} for \gls{cscg} signaling with various receivers and numbers $S$ of \gls{sic} stages. Plot (a) is for the \gls{cpan} channel and plot (b) is for the nonlinear fiber-optic channel. The solid black curves show a capacity upper bound \cite{Kramer:15:UpperBound,Yousefi:15:UpperBound}.}
    \label{fig:results_gaussian}
\end{figure}

%% file: plots/air_v_stages.tex
\begin{figure}[t!]
    \centering
    \begin{tikzpicture}
        \begin{axis}[%
            width=6cm,
            height=3cm,
            at={(0,0)},
            scale only axis,
            xmin=0,
            xmax=6,
            ymin=8.2,
            ymax=8.9,
            grid = major,
            xlabel = {$\nstages$},
            ylabel = {AIR [bpcu]},
            xtick = {0,1,2,3,4,5,6},
            xticklabels = {1,2,4,8,16,32,64},
            axis background/.style={fill=white},            
            legend pos=south east,
            legend style={nodes={scale=0.8, transform shape}, fill=white},
            ]
            \addplot [color=matlab1, dashed, forget plot, mark = x]
              table[]{plots/air_v_stages-1.tsv};
            \addplot [color=matlab1, mark = x]
              table[]{plots/air_v_stages-2.tsv};
            \addlegendentry{\gls{sic}};
            \addplot [color=black]
              table[]{plots/air_v_stages-3.tsv};
            \addlegendentry{\gls{jdd}};
            \addplot [color=black, dashed, forget plot]
              table[]{plots/air_v_stages-4.tsv};
        \end{axis}
    \end{tikzpicture}
    \caption{Maximum \glspl{air} vs. the number of stages for the \gls{cpan} channel (dashed lines) and the nonlinear fiber-optic channel (solid lines).}
    \label{fig:air_v_stages}
\end{figure}
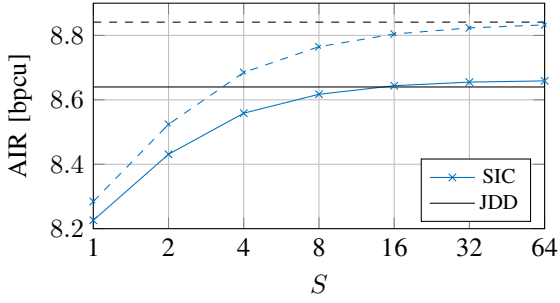

%% file: SIC_Ring.tex
\section{SIC for Ring Constellations}
\label{sec:ring}
Gaussian signaling is suitable for performance analysis, but practical transmitters use discrete constellations. The receiver presented in Sec.~\ref{sec:gaussian} relies on continuously uniform phase signaling, and we next study ring constellations with discrete amplitude and continuous phase. An extension to fully discrete amplitude and phase is given in \cite{Jaeger:24:ecoc}.

Consider ring constellations with independent amplitude $R_i$ and phase $\Gamma_i$. The alphabets of the amplitude and phase are $\mathcal{R}=\{\tilde{r}_1,\ldots,\tilde{r}_\nrings\}$ and $[-\pi,\pi)$, respectively. We use the distribution and density
\begin{align}
    P(\tilde{r}_i) = w_i, \quad
    p(\gamma_i) = \frac{1}{2\pi}, \text{ for } \gamma_i\in[-\pi,\pi).
\end{align}
Like \gls{cscg} inputs, this ring constellations is circularly symmetric, and we have (see \eqref{equ:thetaB_p})
\begin{equation}
    \etaB_{\theta_i'}(\theta) = \normB{\theta_i'}\int_\mathbb{C} p(x)q(y_i|x,\theta)\,\mathrm{d}x=\const
\end{equation}

Motivated by probabilistic shaping for the \gls{cscg} inputs, we use equidistant rings with $\tilde{r}_\ell = \ell\cdot\Delta r$ and ring probabilities
\begin{equation}
    w_\ell = \frac{\tilde{r}_\ell\exp\left(-\frac{\tilde{r}_\ell^2}{\sigma_x^2}\right)}{\sum_{m=1}^\nrings \tilde{r}_m\exp\left(-\frac{\tilde{r}_m^2}{\sigma_x^2}\right)}.
\end{equation}
that model a Rayleigh distribution with variance $\sigma_x^2$. We name this constellation \gls{urr}. The transmit power is
\begin{equation}
\label{equ:ring_power}
    \mathrm{E}[|X|^2] = \sigma_x^2 = \Delta r^2\frac{\sum_{\ell=1}^\nrings \ell^3\exp\left(-\frac{\ell^2\Delta r^2}{\sigma_x^2}\right)}{\sum_{\ell=1}^\nrings \ell\exp\left(-\frac{\ell^2\Delta r^2}{\sigma_x^2}\right)}
\end{equation}
and we set $\sigma_x^2=\ptx$. We numerically determine the $\Delta r$ that satisfies the power constraint. 

\gls{urr} constellations approximate a \gls{cscg} for large $\nrings$. \figref{fig:AIR_AWGN_Ring_Gaussian} shows the \glspl{air} for memoryless \gls{awgn} channels with $\sigma_n^2=2.95\cdot 10^{-7}$, which is approximately the \gls{ase} noise variance for the parameters in \tabref{tab:sim_parameters}. The horizontal line indicates the largest \gls{air} of 2  \gls{sic}-stages in \figref{fig:results_gaussian_b}, which is the peak rate
we wish to achieve using \gls{urr} constellations and 2 \gls{sic}-stages. Thirty-two rings are needed to prevent significant deviation from Gaussian inputs at the target \gls{air}.

\input{plots/Ring_AWGN}

Finally, one may instead use geometric shaping for the ring amplitudes, i.e., uniformly distributed amplitudes with square-root logarithmic spacing; see~\cite[Appendix~A.A]{Essiambre:10:Capacity}.

\subsection{Mutual Information Estimation}

Suppose $\vek{X}$ has independent amplitudes $\vek{R}$ and phases $\vek{\Gamma}$. By the chain rule of \gls{mi}, we have
\begin{equation}
    I(\vek{X};\vek{Y})
    = I(\vek{R},\vek{\Gamma};\vek{Y})
    = I(\vek{R};\vek{Y}) + I(\vek{\Gamma};\vek{Y}|\vek{R}).
    \label{eq:I(R,Theta;Y)}
\end{equation}
Consider \eqref{equ:splitting_x} and write
\begin{align}
\label{equ:ring_partitioning} 
    & \vek{\gamma} = [\alpha_1,\beta_1,\alpha_2,\beta_2,\ldots,\alpha_{n/2},\beta_{n/2}] \\
    & a_i = r_{2i-1}\exp(\imag\alpha_i), \quad
    b_i = r_{2i}\exp(\imag\beta_i)
\end{align}
so the rightmost term in \eqref{eq:I(R,Theta;Y)} is
\begin{equation}
\label{equ:mi_sic_ring}
    I(\vek{\Gamma};\vek{Y}|\vek{R})
    = I(\vek{\alpha};\vek{Y}|\vek{R}) + I(\vek{\beta};\vek{Y}|\vek{R},\vek{\alpha}).
\end{equation}
We divide only the phase vector into components related to $\vek{a}$ and $\vek{b}$ because, as we show below, the amplitudes can be detected and decoded separately.

A \gls{gmi} for the amplitudes is
\begin{align}    
    \label{equ:mi_r}I_R(\vek{R};\vek{Y}) &:= \sum_{i=1}^\nsym H(R_i)-H(R_i|\vek{Y})\\
    \nonumber&\leq  I(\vek{R};\vek{Y})
\end{align}
and \glspl{gmi} for the phases are
\begin{align}    
    \label{equ:mi_alpha}I_1(\vek{\alpha};\vek{Y}|\vek{R}) &:= \sum_{i=1}^{\nsym/2} h(\alpha_i)-h(\alpha_i|\vek{Y},\vek{R})\\ 
    \nonumber&\leq I(\vek{\alpha};\vek{Y}|\vek{R})\\
    \label{equ:mi_beta}I_2(\vek{\beta};\vek{Y}|\vek{R},\vek{\alpha}) &:= \sum_{i=1}^{\nsym/2} h(\beta_i)-h(\beta_i|\vek{Y},\vek{R},\vek{\alpha})\\
    \nonumber & \leq I(\vek{\beta};\vek{Y}|\vek{R},\vek{\alpha}).
\end{align}

An \gls{air} for \gls{sic} is thus
\begin{align}  
        & \frac{1}{\nsym}I_\mathrm{sic}(\vek{X};\vek{Y}) \nonumber \\
        &:= \frac{1}{\nsym}\Big(I_R(\vek{R};\vek{Y}) + I_1(\vek{\alpha};\vek{Y}|\vek{R}) + I_2(\vek{\beta};\vek{Y}|\vek{R},\vek{\alpha})\Big) \nonumber \\
        & \le \frac{1}{\nsym}I(\vek{X};\vek{Y}).
\end{align}

Similar to \eqref{equ:mismatched_prob}, consider
\begin{align}  
    & q(\vek{r},\vek{\gamma},\vek{y},\vek{\theta})
    =\prod_{i=1}^\nsym P(r_i)p(\gamma_i)q(\theta_i|\theta_{i-1})q(y_i|r_i,\gamma_i,\theta_i) .
\end{align}
We again discard dependencies for the sake of clarity. The receiver wishes to calculate
\begin{align}
    q(r_i|\vek{y}) &= \frac{1}{c_3}\int\displaylimits_{\mathbb{R}^n}\int\displaylimits_{\mathcal{\vek{\Pi}}}\sum_{\vek{r}\in\vek{\mathcal{R}}^{\backslash\{i\}}} q(\vek{r},\vek{\gamma},\vek{y},\vek{\theta})\, \mathrm{d}\vek{\gamma}\,\mathrm{d}\vek{\theta}\\
    q(\alpha_i|\vek{y},\vek{r}) &= \frac{1}{c_4}\int\displaylimits_{\mathbb{R}^n}\int\displaylimits_{\mathcal{\vek{\Pi}}^{\backslash\{i\}}} q(\vek{r},\vek{\gamma},\vek{y},\vek{\theta}) \,\mathrm{d}\vek{\gamma}\,\mathrm{d}\vek{\theta}\\
    q(\beta_i|\vek{y},\vek{r},\vek{\alpha}) &= \frac{1}{c_5}\int\displaylimits_{\mathbb{R}^n}\int\displaylimits_{\mathcal{\vek{\Pi}}_{\vek{\alpha}}^{\backslash\{i\}}} q(\vek{r},\vek{\gamma},\vek{y},\vek{\theta}) \,\mathrm{d}\vek{\gamma}\,\mathrm{d}\vek{\theta}
\end{align}
where
\begin{align}
    \vek{\mathcal{R}}^{\backslash\{i\}}&=\{\vek{r}'\in\mathcal{R}^\nsym:r'_i=r_i\}\\
    \mathcal{\vek{\Pi}} &= [-\pi,\pi)^\nsym\\
    \mathcal{\vek{\Pi}}^{\backslash\{i\}}&=\{\vek{\gamma}\in[-\pi,\pi)^\nsym:\gamma_{2i-1}=\alpha_i\}\\
    \mathcal{\vek{\Pi}}_{\vek{\alpha}}^{\backslash \{i\}}&=\{\vek{\gamma}\in[-\pi,\pi)^\nsym:\vek{\alpha}(\vek{\gamma})=\vek{\alpha}\land \gamma_{2i}=\beta_i\}.
\end{align}
with $\vek{\alpha}(\vek{\gamma})=[\gamma_1,\gamma_3,\ldots,\gamma_{n-1}]$. As before, we marginalize $q(\vek{r},\vek{\gamma},\vek{y},\vek{\theta})$ where the variables subject to marginalization depend on the \gls{sic}-stage.

\subsection{Computing the Marginal Distributions}
\subsubsection{Amplitude Detection}
\begin{figure}[t!]
    \centering
    \begin{tikzpicture}
    \node[draw, shape = rectangle, very thick] (ptheta1) at (0,0) {\small$q(\theta_i|\theta_{i-1})$};
    \node[draw, shape = rectangle, right = 1.8cm of ptheta1, very thick] (equ1) {\large$=$};
    \node[draw, shape = rectangle, right = 1.8cm of equ1, very thick] (ptheta2){\small$q(\theta_{i+1}|\theta_i)$};
    
    \node[draw, shape = rectangle, below = 2cm of equ1, very thick] (qy1) {\small $q(y_i|r_i,\gamma_i,\theta_i)$};
    \node[below = 2cm of qy1] (midpoint) {};
    \node[draw, shape = rectangle, right = 2cm of midpoint, very thick] (pgamma1) {\small$p(\gamma_i)$};
    \node[draw, shape = rectangle, left = 2cm of midpoint, very thick] (pr1) {\small$P(r_i)$};

    \draw[->, very thick] (ptheta1)--(equ1) node[midway,fill=white] {$\theta_i$};
    \draw[->, very thick] (equ1)--(ptheta2) node[midway,fill=white] {$\theta_i''$};
    
    \draw[->, very thick] (equ1)--(qy1) node[midway,fill=white] {$\theta_i'$};
    \draw[->, very thick] (pgamma1)--(qy1) node[midway,fill=white] {$\gamma_i$};
    \draw[->, very thick] (pr1)--(qy1) node[midway,fill=white] {$r_i$};

    \draw[msgarrow] (ptheta1) edge[bend left, ->] node[above] {$\etaF_{\theta_i}$} (equ1);
    \draw[msgarrow] (equ1) edge[bend left, ->] node[above] {$\etaF_{\theta_i''}$} (ptheta2);
    
    \draw[msgarrow] (equ1) edge[bend right, <-] node[below] {$\etaB_{\theta_i''}$} (ptheta2);
    \draw[msgarrow] (ptheta1) edge[bend right, <-] node[below] {$\etaB_{\theta_i}$} (equ1);
    
    \node[left = 0.1cm of ptheta1] (leftdots) {$\ldots$};
    \node[right = 0.1cm of ptheta2] (rightdots) {$\ldots$};
    \draw[very thick] (ptheta2)--(rightdots);
    \draw[very thick] (leftdots)--(ptheta1);
    
    \draw[msgarrow] (pgamma1) edge[bend left, ->] node[left] {$\etaF_{\gamma_i}$} (qy1);    
    \draw[msgarrow] (pgamma1) edge[bend right, <-] node[right] {$\etaB_{\gamma_i}$} (qy1);
    \draw[msgarrow] (pr1) edge[bend left, ->] node[left] {$\etaF_{r_i}$} (qy1);    
    \draw[msgarrow] (pr1) edge[bend right, <-] node[right] {$\etaB_{r_i}$} (qy1);
    
    \draw[msgarrow] (qy1) edge[bend left, ->] node[left, pos=1/4] {$\etaB_{\theta_i'}$} (equ1);
    
    \draw[msgarrow] (qy1) edge[bend right, <-] node[right, pos=1/4] {$\etaF_{\theta_i'}$} (equ1);
    
    \end{tikzpicture}
    \caption{Branch of the non-decoded stage with non-decoded amplitude.}
    \label{fig:first_stage_ring}
\end{figure}
The graph to detect the amplitudes $\vek{r}$ has branches shown in \figref{fig:first_stage_ring}. Using 
\begin{equation}
\label{equ:integral_cscg}
    \int_{-\pi}^\pi q(y_i|r_i,\gamma,\theta_i)\,\mathrm{d}\gamma = \frac{2}{\sigma_n^2}
    e^{-\left(|y_i|^2+r_i^2\right)/\sigma_n^2}
    I_0\left(\frac{2|y_i|r_i}{\sigma_n^2}\right)
\end{equation}
and $p(\gamma)=\frac{1}{2\pi}$, one can again show that
\begin{align}   
    \etaB_{\theta_i'}(\theta)&=\normB{\theta_i'}\sum_{r\in\mathcal{R}}\int_{-\pi}^\pi p(\gamma)P(r)q(y_i|r,\gamma,\theta)\,\mathrm{d}\gamma \nonumber \\
    &= \const
\end{align}
As in \eqref{equ:cscg_etaF_thetai_first_stage}, we obtain
\begin{equation}
    \etaF_{\theta_i'}(\theta) = q_{\theta}(\theta)
\end{equation}
and
\begin{align}
    \etaB_{r_i}(r) &= \normB{r_i}\int_\mathbb{R} \etaF_{\theta_i'}(\theta)\int_{-\pi}^\pi p(\gamma)q(y_i|r,\gamma,\theta)\,\mathrm{d}\gamma\,\mathrm{d}\theta \nonumber \\
    &\propto \exp\left(-\frac{r^2}{\sigma_n^2}\right)I_0\left(\frac{2|y_i|r}{\sigma_n^2}\right).
        \label{equ:etaBri}
\end{align}
With this, upon receiving $y_i$, one can compute
\begin{equation}
\label{equ:q_r}
    q(r_i|\vek{y}) = \frac{P(r_i)\etaB_{r_i}(r_i)}{\sum_{\tilde{r}\in\mathcal{R}}P(\tilde{r})\etaB_{r_i}(\tilde{r})}.
\end{equation}
Note that the computations for different $i$ may run in parallel, and a memoryless receiver can be used as $q(r_i|\vek{y})=q(r_i|y_i)$.

We now investigate \gls{sic} with two stages for amplitude detection. In the second stage, branches of the type shown in \figref{fig:first_stage_ring} and \figref{fig:second_stage_ring} alternate. For odd $i$, we have branches of the form shown in \figref{fig:second_stage_ring} and
\begin{align}
    \etaB_{\theta_i'}(\theta) &= \normB{\theta_i'}\int_{-\pi}^\pi p(\gamma)q(y_i|r,\gamma,\theta)\mathrm\,{d}\gamma \nonumber \\
    &= \const
\end{align}
where we used \eqref{equ:integral_cscg}. Following the same steps as before, we recover \eqref{equ:q_r}. Therefore, the receiver does not use the entries of $\vek{r}$ decoded in the first stage. We can hence use \eqref{equ:q_r} to detect all elements in $\vek{r}$ and achieve no gain using \gls{sic}. This motivates using the partition of \eqref{equ:splitting_x} with \eqref{equ:ring_partitioning}.

\begin{figure}[t!]
    \centering
    \begin{tikzpicture}
    \node[draw, shape = rectangle, very thick] (ptheta1) at (0,0) {\small$q(\theta_i|\theta_{i-1})$};
    \node[draw, shape = rectangle, right = 1.8cm of ptheta1, very thick] (equ1) {\large$=$};
    \node[draw, shape = rectangle, right = 1.8cm of equ1, very thick] (ptheta2){\small$q(\theta_{i+1}|\theta_i)$};
    
    \node[draw, shape = rectangle, below = 2cm of equ1, very thick] (qy1) {\small $P(r_i)q(y_i|r_i,\gamma_i,\theta_i)$};
    \node[draw, shape = rectangle, below = 2cm of qy1, very thick] (px1) {\small$p(\gamma_i)$};

    \draw[->, very thick] (ptheta1)--(equ1) node[midway,fill=white] {$\theta_i$};
    \draw[->, very thick] (equ1)--(ptheta2) node[midway,fill=white] {$\theta_i''$};
    
    \draw[->, very thick] (equ1)--(qy1) node[midway,fill=white] {$\theta_i'$};
    \draw[->, very thick] (px1)--(qy1) node[midway,fill=white] {$\gamma_i$};

    \draw[msgarrow] (ptheta1) edge[bend left, ->] node[above] {$\etaF_{\theta_i}$} (equ1);
    \draw[msgarrow] (equ1) edge[bend left, ->] node[above] {$\etaF_{\theta_i''}$} (ptheta2);
    
    \draw[msgarrow] (equ1) edge[bend right, <-] node[below] {$\etaB_{\theta_i''}$} (ptheta2);
    \draw[msgarrow] (ptheta1) edge[bend right, <-] node[below] {$\etaB_{\theta_i}$} (equ1);
    
    \node[left = 0.1cm of ptheta1] (leftdots) {$\ldots$};
    \node[right = 0.1cm of ptheta2] (rightdots) {$\ldots$};
    \draw[very thick] (ptheta2)--(rightdots);
    \draw[very thick] (leftdots)--(ptheta1);
    
    \draw[msgarrow] (px1) edge[bend left, ->] node[left] {$\etaF_{\gamma_i}$} (qy1);
    
    \draw[msgarrow] (px1) edge[bend right, <-] node[right] {$\etaB_{\gamma_i}$} (qy1);
    
    \draw[msgarrow] (qy1) edge[bend left, ->] node[left, pos=1/4] {$\etaB_{\theta_i'}$} (equ1);
    
    \draw[msgarrow] (qy1) edge[bend right, <-] node[right, pos=1/4] {$\etaF_{\theta_i'}$} (equ1);
    
    \end{tikzpicture}
    \caption{Branch of the non-decoded stage with decoded amplitude.}
    \label{fig:second_stage_ring}
\end{figure}

\subsubsection{Phase Detection, First Stage}
The graph is a concatenation of branches of the form shown in \figref{fig:second_stage_ring}. Using \eqref{equ:integral_cscg}, we again have $\etaB_{\theta_i'}(\theta)=\const$ and $\etaF_{\theta_i'}(\theta)= q(\theta)$. Similar to \eqref{equ:periodicity_argument}, we obtain 
\begin{align}   
    &q(y_i|r_i,\gamma_i,\theta) \approx \nonumber \\ 
    & \frac{1}{\sqrt{|y_i|r_i}}\normpdf{|y_i|}{r_i}{\frac{\sigma_n^2}{2}}\normpdf{\theta}{\modpi{\anglenobr{y_i}-\gamma_i}}{\frac{\sigma_n^2}{2|y_i|r_i}}.
\end{align}

Using $\etaF_{\theta_i'}(\theta)=\normpdf{\theta}{0}{\thetavar}$, we thus have
\begin{align}
    q(\gamma_i|\vek{y},\vek{r}) &= \frac{1}{c_6}\etaF_{\gamma_i}(\gamma_i)\int_\mathbb{R} \etaF_{\theta_i'}(\theta)q(y_i|r_i,\gamma_i,\theta)\mathrm{d}\theta \nonumber \\
    &\approx\normpdf{\modpi{\anglenobr{y_i}-\gamma_i}}{0}{\thetavar+\frac{\sigma_n^2}{2|y_i|r_i}}.
    \label{equ:phase_detection}
\end{align}
The scaling constant ensures \eqref{equ:phase_detection} has unit integral over the support of $\gamma_i$. However, as the tails decay rapidly, this constant is larger than but very close to 1 and may be omitted. 

\subsubsection{Phase Detection, Second Stage}
Branches of the form shown in \figref{fig:second_stage} for odd $i$ and \figref{fig:second_stage_ring} for even $i$ alternate. In the former case, we use the approximation (\ref{equ:etaBthetaip}), while in the latter case $\etaB_{\theta_i'}(\theta)$ is constant in $\theta$. With the same steps as before, we obtain
\begin{equation}
    \etaF_{\theta_i'}(\theta)\approx\normpdf{\theta}{\muF_{\theta_i'}}{\sigmaF_{\theta_i'}^2}
\end{equation}
where \eqref{equ:muFthetap} and \eqref{equ:sigmaFtheap} give the expressions for $\muF_{\theta_i'}$ and $\sigmaF_{\theta_i'}^2$. Similar to \eqref{equ:phase_detection}, we now have
\begin{equation}
\label{equ:phase_detection_second_stage} q(\gamma_i|\vek{y},\vek{r},\vek{\alpha}) = \normpdf{\modpi{\anglenobr{y_i}-\gamma_i-\muF_{\theta_i'}}}{0}{\sigmaF_{\theta_i'}^2+\frac{\sigma_n^2}{2|y_i|r_i}}
\end{equation}
where we omitted the normalization, as discussed above.

\subsection{Extension to $\nstages$ \gls{sic}-Stages}
Extending the algorithm to $\nstages$ stages is straightforward. We first decode $\vek{r}$ using \eqref{equ:q_r} and the first stage of $\vek{\gamma}$ using \eqref{equ:phase_detection}. For stage $s$, we reuse algorithm \ref{alg:gaussian} to obtain $\etaFvec_{\theta_i'}=\left[\muF_{\theta'_i},\sigmaF_{\theta_i'}^2\right]$ and then $q(\gamma_i|\vek{y},\vek{r},\vek{\gamma}^{(1)},\ldots,\vek{\gamma}^{(s-1)})$ for $i\in \{s,s+\nstages,\ldots\}$, as indicated by \eqref{equ:phase_detection_second_stage}.
    
\subsection{Lower Bound on Mutual Information}
We use the same approach as in \secref{subsec:lower_bound_mi} and define
\begin{align}
    I_{R,q}(\vek{R};\vek{Y}) &:= \sum_{i=1}^\nsym H(R_i)-H_q(R_i|Y_i) \\
    & \le I_{R}(\vek{R};\vek{Y}) \nonumber
\end{align}
for the amplitudes and
\begin{align}
    I_{1,q}(\vek{\alpha};\vek{Y}|\vek{R}) &:=\sum_{i=1}^{\nsym/2} h(\alpha_i)-h_q(\alpha_i|\vek{Y},\vek{R})\\&\nonumber\leq I_{1}(\vek{\alpha};\vek{Y}|\vek{R})\\
    I_{2,q}(\vek{\beta};\vek{Y}|\vek{R},\vek{\alpha}) &:= \sum_{i=1}^{\nsym/2} h(\beta_i)-h_q(\beta_i|\vek{Y},\vek{R},\vek{\alpha})\\&\nonumber\leq I_{2}(\vek{\beta};\vek{Y}|\vek{R},\vek{\alpha})
\end{align}
for the phases. Note that, for all $i$, we have
\begin{align}
    H(R_i) = -\sum_{\ell=1}^\nrings w_\ell\log w_\ell, \quad
    h(\alpha_i) =h(\beta_i)=\log 2\pi .
\end{align}
The surrogate channel conditional (differential) entropies can be approximated by simulation as in \secref{subsec:lower_bound_mi}.

\subsection{Simulation Results}
\input{plots/Ring_air}
\input{plots/Ring_AIR_Phase_Amplitude}

\figref{fig:results_ring_2_stages} shows the \glspl{air} for $S=2$ and transmission over the nonlinear fiber-optic channel. The \gls{air} of \gls{cscg} modulation is plotted in dashed black for reference. As expected from \figref{fig:AIR_AWGN_Ring_Gaussian}, the \gls{air} increases with the number of rings and saturates at 32 rings.
The phase noise variance depends on the amplitude statistics. For example, $M$-PSK or ring constellations with one ring cause little phase noise, whereas Gaussian modulation causes significant phase noise \cite{Dar:16:Collision}. Thus, we have a tradeoff: increasing the number of rings increases the amplitude channel's rate but also the phase noise variance. The left plot in \figref{fig:results_ring_phase_amplitude} shows that for two \gls{sic}-stages, the \gls{air} of the phase channel decreases with an increasing number of rings. The right side shows that the \gls{air} of the amplitude channel increases by a larger amount, and hence, the overall \gls{air} increases for an increasing number of rings. 

\input{plots/air_v_stages_urr}
\input{plots/ring_32_air}

\figref{fig:air_v_stages_urr} plots the maximum \gls{air} as a function of the number $\nstages$ of stages for ring and \gls{cscg} signaling, as well as \gls{cscg} signaling with a \gls{jdd} detector; cf.~\figref{fig:air_v_stages}. Observe that $32$ rings performs similarly to \gls{cscg} modulation. Increasing the number of \gls{sic}-stages beyond $16$ does not improve the \gls{air}. 
\figref{fig:results_ring_32_rings} shows the rates as a function of the number of \gls{sic}-stages for 32 rings. This is similar to the results in \figref{fig:results_gaussian}.

%% file: plots/Ring_AWGN.tex
\begin{figure}[t!]
    \centering
    \begin{tikzpicture}
        \begin{axis}[%
            width=5cm,
            height=3cm,
            scale only axis,
            xmin=10,
            xmax=36,
            xtick = {10,20,30},
            ymin=4,
            ymax=11.5,
            ylabel = {AIR [bpcu]},
            xlabel = {SNR [dB]},
            axis x line*=top,
            axis background/.style={fill=white}
            ]
            \addplot [color=matlabBlue, forget plot]
              table[]{plots/Ring_AWGN-2.tsv};
            \addplot [color=matlabBlue, forget plot]
              table[]{plots/Ring_AWGN-3.tsv};
            \addplot [color=matlabBlue, forget plot]
              table[]{plots/Ring_AWGN-4.tsv};
            \addplot [color=matlabBlue, forget plot]
              table[]{plots/Ring_AWGN-5.tsv};
            \addplot [color=white!15!black, forget plot]
              table[]{plots/Ring_AWGN-6.tsv};
            \addplot [color=matlabRed, forget plot]
              table[]{plots/Ring_AWGN-1.tsv};
        \end{axis}
        \begin{axis}[%
            width=5cm,
            height=3cm,
            scale only axis,
            xmin=-25.3004,
            xmax=0.6996,
            xtick = {-15,-5,5},
            ymin=4,
            ymax=11.5,
            ylabel = {AIR [bpcu]},
            xlabel = {$\ptx$ [dBm]},
            axis x line* = bottom,
            grid = {major},
            legend pos=north west,
            legend style={nodes={scale=0.8, transform shape}, fill=white},
            ]
            \addplot [color=matlabRed, draw = none]
              table[]{plots/Ring_AWGN-1.tsv};
            \addlegendentry{Gaussian};
            \addplot [color=matlabBlue, draw = none]
              table[]{plots/Ring_AWGN-2.tsv};
            \addlegendentry{Ring};
            \addplot [draw=none, forget plot]
              table[]{plots/Ring_AWGN_Ptx-1.tsv};
            \node[fill=white,text=matlabBlue, inner sep = 0pt] at (-0.5,8.32) {\small$4$};
            \node[fill=white,text=matlabBlue, inner sep = 0pt] at (-0.5,9.1) {\small$8$};
            \node[fill=white,text=matlabBlue, inner sep = 0pt] at (-0.5,9.92) {\small$16$};
            \node[fill=white,text=matlabBlue, inner sep = 0pt] at (-0.5,10.78) {\small$32$};
        \end{axis}
    \end{tikzpicture}
    \caption{\gls{air} of Gaussian and ring constellations for memoryless \gls{awgn} channels with noise variance $2.95\cdot10^{-7}$. Horizontal line indicates the highest \gls{air} for $2$ \gls{sic}-stages in \figref{fig:results_gaussian_b}. Numbers on lines indicate the number of rings.}
    \label{fig:AIR_AWGN_Ring_Gaussian}
\end{figure}

%% file: plots/Ring_air.tex
\begin{figure}[t!]
    \centering
    \begin{tikzpicture}    
        \begin{axis}[%
        width=6cm,
        height=4cm,
        at={(0,0)},
        scale only axis,
        xmin=-14,
        xmax=-3,
        ymin=6.5,
        ymax=9.1,
        xlabel = {$\ptx$ [dBm]},
        ylabel = {AIR [bpcu]},
        grid = major,
        axis background/.style={fill=white},
        legend pos=north west,
        legend style={nodes={scale=0.7, transform shape}},
        legend columns = 3
        ]
            \addplot [color=black, dashed, style=thick, mark = *, mark size = .3pt]
              table[]{plots/Ring_AIR_2_Stages-1.tsv};
              \addlegendentry{Gaussian};
            \addplot [color=matlab1, style=thick, mark = *]
              table[]{plots/Ring_AIR_2_Stages-2.tsv};
              \addlegendentry{4 Rings};
            \addplot [color=matlab2, style=thick, mark = square]
              table[]{plots/Ring_AIR_2_Stages-3.tsv};
              \addlegendentry{8 Rings};
            \addplot [color=matlab3, style=thick, mark = diamond]
              table[]{plots/Ring_AIR_2_Stages-4.tsv};
              \addlegendentry{16 Rings};
            \addplot [color=matlab4, style=thick, mark = o]
              table[]{plots/Ring_AIR_2_Stages-5.tsv};
              \addlegendentry{32 Rings};
            \addplot [color=matlab5, style=thick, mark = x]
              table[]{plots/Ring_AIR_2_Stages-6.tsv};
              \addlegendentry{64 Rings};
            \addplot [color=black, dashed, forget plot, style=thick, mark = *, mark size = .3pt]
            table[]{plots/Ring_AIR_2_Stages-1.tsv};
        \end{axis}
    \end{tikzpicture}%
    \caption{\gls{air} for 2 \gls{sic}-stages with Gaussian modulation and ring constellations. Transmission is over the nonlinear fiber-optic channel.}
    \label{fig:results_ring_2_stages}
\end{figure}
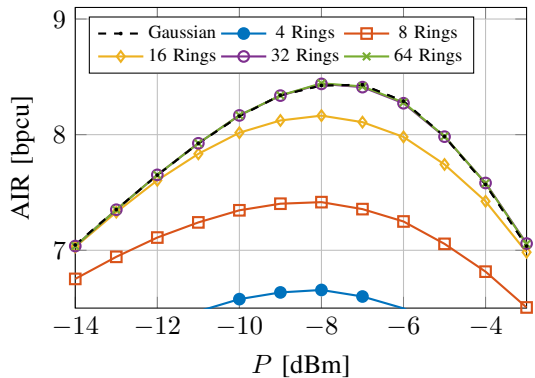

%% file: plots/Ring_AIR_Phase_Amplitude.tex
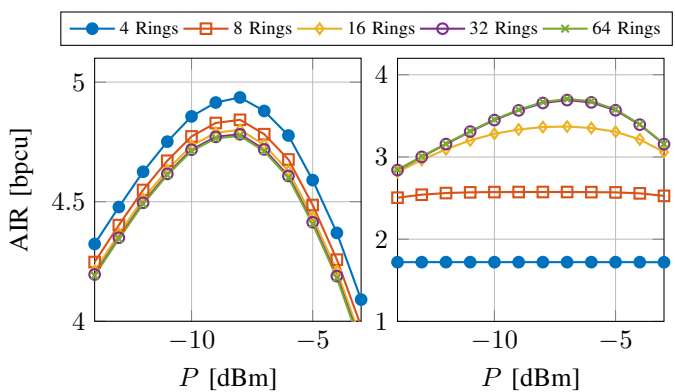
\begin{figure}[t!]
    \centering
    \begin{tikzpicture}    
        \begin{axis}[%
        width=\linewidth/2.5,
        height=3.5cm,
        at={(0,0)},
        scale only axis,
        xmin=-14,
        xmax=-3,
        ymin=4,
        ymax=5.1,
        xlabel = {$\ptx$ [dBm]},
        ylabel = {AIR [bpcu]},
        grid = major,
        axis background/.style={fill=white},
        legend style={at={(\linewidth/2.45,4cm)},anchor=south,nodes={scale=0.7, transform shape}},
        legend columns = 5
        ]
            \addplot [color=matlab1, style=thick, mark = *]
              table[]{plots/Ring_AIR_2_Stages_Phase-1.tsv};
              \addlegendentry{4 Rings};
            \addplot [color=matlab2, style=thick, mark = square]
              table[]{plots/Ring_AIR_2_Stages_Phase-2.tsv};
              \addlegendentry{8 Rings};
            \addplot [color=matlab3, style=thick, mark = diamond]
              table[]{plots/Ring_AIR_2_Stages_Phase-3.tsv};
              \addlegendentry{16 Rings};
            \addplot [color=matlab4, style=thick, mark = o]
              table[]{plots/Ring_AIR_2_Stages_Phase-4.tsv};
              \addlegendentry{32 Rings};
            \addplot [color=matlab5, style=thick, mark = x]
              table[]{plots/Ring_AIR_2_Stages_Phase-5.tsv};
              \addlegendentry{64 Rings};
        \end{axis}
        \begin{axis}[%
        width=\linewidth/2.5,
        height=3.5cm,
        at={(\linewidth/2.2,0)},
        scale only axis,
        xmin=-14,
        xmax=-3,
        ymin=1,
        ymax=4.2,
        xlabel = {$\ptx$ [dBm]},
        grid = major,
        axis background/.style={fill=white},
        ]
            \addplot [color=matlab1, forget plot, style=thick, mark = *]
              table[]{plots/Ring_AIR_2_Stages_Amplitude-1.tsv};
            \addplot [color=matlab2, forget plot, style=thick, mark = square]
              table[]{plots/Ring_AIR_2_Stages_Amplitude-2.tsv};
            \addplot [color=matlab3, forget plot, style=thick, mark = diamond]
              table[]{plots/Ring_AIR_2_Stages_Amplitude-3.tsv};
            \addplot [color=matlab4, forget plot, style=thick, mark = o]
              table[]{plots/Ring_AIR_2_Stages_Amplitude-4.tsv};
            \addplot [color=matlab5, forget plot, style=thick, mark = x]
              table[]{plots/Ring_AIR_2_Stages_Amplitude-5.tsv};
        \end{axis}
    \end{tikzpicture}%
    \caption{\glspl{air} of phase (left) and amplitude (right) channel for 2 \gls{sic}-stages and ring constellations. Transmission is over the nonlinear fiber-optic channel.}
    \label{fig:results_ring_phase_amplitude}
\end{figure}

%% file: plots/air_v_stages_urr.tex
\begin{figure}[t!]
    \centering
    \begin{tikzpicture}
        \begin{axis}[%
            width=6cm,
            height=4cm,
            at={(0,0)},
            scale only axis,
            xmin=1,
            xmax=5,
            ymin=8,
            ymax=9,
            grid = major,
            xlabel = {$\nstages$},
            ylabel = {AIR [bpcu]},
            xtick = {1,2,3,4,5},
            xticklabels = {1,2,4,8,16,32},
            axis background/.style={fill=white},  
            legend style={nodes={scale=0.7, transform shape}, anchor = north, at = {(0.5,1)}},
            legend columns = 2
            ]
            \addplot [color=matlab3, mark = diamond]
              table[]{plots/air_v_stages_urr-3.tsv};
            \addlegendentry{16 Rings};
            \addplot [color=matlab4, mark = o]
              table[]{plots/air_v_stages_urr-4.tsv};
            \addlegendentry{32 Rings};

            \addplot [color=black, mark = x, dashed]
              table[]{plots/air_v_stages_urr-6.tsv};
            \addlegendentry{Gaussian SIC};
            \addplot [color=black]
              table[]{plots/air_v_stages_urr-7.tsv};
              \addlegendentry{Gaussian JDD};
        \end{axis}
    \end{tikzpicture}
    \caption{\glspl{air} vs. $\nstages$ for ring and Gaussian constellations. Transmission is over the nonlinear fiber-optic channel.}
    \label{fig:air_v_stages_urr}
\end{figure}
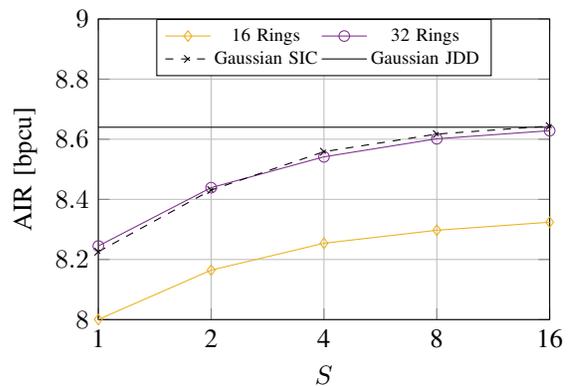

%% file: plots/ring_32_air.tex
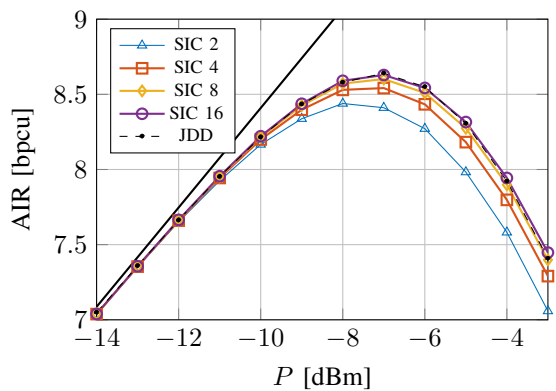
\begin{figure}[t!]
    \centering
    \begin{tikzpicture}    
        \begin{axis}[%
        width=6cm,
        height=4cm,
        at={(0,0)},
        scale only axis,
        xmin=-14,
        xmax=-3,
        ymin=7,
        ymax=9,
        xlabel = {$\ptx$ [dBm]},
        ylabel = {AIR [bpcu]},
        grid = major,
        axis background/.style={fill=white},
        legend pos=north west,
        legend style={nodes={scale=0.7, transform shape}}
        ]
            \addplot [color=matlab1, mark = triangle]
              table[]{plots/Ring_AIR_32_Rings-1.tsv};
              \addlegendentry{SIC 2};
            \addplot [color=matlab2, style=thick, mark = square]
              table[]{plots/Ring_AIR_32_Rings-2.tsv};
              \addlegendentry{SIC 4};
            \addplot [color=matlab3, style=thick, mark = diamond]
              table[]{plots/Ring_AIR_32_Rings-3.tsv};
              \addlegendentry{SIC 8};
            \addplot [color=matlab4, style=thick, mark = o]
              table[]{plots/Ring_AIR_32_Rings-4.tsv};
              \addlegendentry{SIC 16};
            \addplot [dashed, color=black, mark = *, mark size = .8pt]
                      table[]{plots/Gaussian_AIR-9.tsv};
              \addlegendentry{JDD};
            \addplot [color=black, forget plot, style=thick]
                  table[]{plots/air_upper_bound_ase-1.tsv};
        \end{axis}
    \end{tikzpicture}%
    \caption{\glspl{air} of ring constellations with 32 rings for different numbers $S$ of \gls{sic}-stages and the \gls{jdd} receiver with Gaussian inputs. The solid black curve is a capacity upper bound \cite{Kramer:15:UpperBound}. Transmission is over the nonlinear fiber-optic channel.}
    \label{fig:results_ring_32_rings}
\end{figure}

%% file: conclusions.tex
\section{Conclusions \& Outlook}
\label{sec:conclusion}
We used \gls{sic}-based receivers to compensate for nonlinearity in optical fiber. The receiver applied the \gls{cpan} model as a surrogate channel and implemented the \gls{spa} with \gls{gmp}. The receiver algorithms for \gls{cscg} modulation and ring constellations give \glspl{air} close to those of \gls{jdd} \cite{Gomez:20:CPAN} for 16 or more \gls{sic}-stages. In contrast to \gls{jdd}, there is a path to implementation with coded modulation for memoryless channels. The ring constellations perform as well as \gls{cscg} modulation for 32 or more rings. We extended the approach to discrete constellations in recent work \cite{Jaeger:24:ecoc}.

For future work, we plan to study multistage encoding, dual-polarization transmission, space-division multiplexing, and lumped amplification instead of \gls{idra}. Another idea is to discard digital backpropagation and use the proposed receiver to compensate for self-phase modulation. 

%% file: appendix.tex
\appendices
\section{Moments of $f_i(\cdot)$ in \eqref{equ:fi}}
\label{app:derivation_f}
For \gls{cscg} inputs, consider the density
\begin{equation}
    f(x) = \frac{1}{c_{f}}p(x)\int_\mathbb{R}\etaF_{\theta'}(\theta)q(y|x,\theta)\,\mathrm{d}\theta
\end{equation}
where
\begin{equation}
    \begin{aligned}    
        c_{f} &= \int_\mathbb{R}\etaF_{\theta'}(\theta)\int_\mathbb{C} p(x)q(y|x,\theta)\,\mathrm{d}x\,\mathrm{d}\theta = q(y)
    \end{aligned}
\end{equation}
with $q(y)=\cscgnormpdf{y}{0}{\sigma_y^2}$. Define
\begin{equation}
    g(x) = \begin{cases}
        x,&\text{for } \mathrm{E}_f[X]\\
        |x|^2,&\text{for } \mathrm{E}_f[|X|^2]\\
        x^2,&\text{for } \mathrm{E}_f[X^2]
    \end{cases}
\end{equation}
so the second-order moments can be calculated with
\begin{equation}
    \begin{aligned}
        &\int_\mathbb{C} g(x)f(x)\mathrm{d}x\\
        &=\underbrace{\int_\mathbb{R}\etaF_{\theta'}(\theta)\e^{-k\imag\theta}\mathrm{d}\theta}_{\displaystyle =:a} \int_\mathbb{C} g(\tilde{x})\underbrace{\frac{p(\tilde{x})q(y|\tilde{x},0)}{q(y)}}_{\displaystyle =:b(\tilde{x})}\mathrm{d}\tilde{x}
    \end{aligned}
    \label{eq:app-moments}
\end{equation}
where $\tilde{x}=x\e^{\imag\theta}$, $k=1$ for $\mathrm{E}_f[X]$, $k=0$ for $\mathrm{E}_f[|X|^2]$, and $k=2$ for $\mathrm{E}_f[X^2]$.
Using $\etaF_{\theta'}(\theta)=\normpdf{\theta}{\muF_{\theta'}}{\sigmaF^2_{\theta'}}$ and completing squares gives
\begin{equation}
    a = \exp\left(-\frac{1}{2}\frac{\muF_{\theta'}^2-(\muF_{\theta'}-k\imag\sigmaF_{\theta'}^2)^2}{\sigmaF_{\theta'}^2}\right).
\end{equation}
Also, $b(\tilde{x})$ in \eqref{eq:app-moments} is a \gls{cscg}
\begin{equation}
    b(\tilde{x}) = \cscgnormpdf{\tilde{x}}{y\frac{\sigma_x^2}{\sigma_y^2}}{\frac{\sigma_x^2\sigma_n^2}{\sigma_y^2}}
\end{equation}
and therefore
\begin{equation}
    \int_\mathbb{C} g(\tilde{x})b(\tilde{x})\,\mathrm{d}\tilde{x}=\begin{cases}
        y\frac{\sigma_x^2}{\sigma_y^2},&\text{for }g(\tilde{x})=\tilde{x}\\
        \frac{\sigma_x^2}{\sigma_y^2}\left(\sigma_n^2+|y|^2\frac{\sigma_x^2}{\sigma_y^2}\right),&\text{for }g(\tilde{x})= |\tilde{x}|^2\\
        y^2\frac{\sigma_x^4}{\sigma_y^4},&\text{for } g(\tilde{x})=\tilde{x}^2
    \end{cases}.
\end{equation}
The moments of $f$ follow directly:
\begin{align}
    \mu_f &= y\frac{\sigma_x^2}{\sigma_y^2}\exp\left(-\frac{1}{2}\frac{\muF_{\theta'}^2-(\muF_{\theta'}-\imag\sigmaF_{\theta'}^2)^2}{\sigmaF_{\theta'}^2}\right)\\
    \sigma_f^2 &= \frac{\sigma_x^2}{\sigma_y^2}\left(\sigma_n^2+|y|^2\frac{\sigma_x^2}{\sigma_y^2}\right)-|\mu_f|^2\\
    p_f^2 &= y^2\frac{\sigma_x^4}{\sigma_y^4}\exp\left(-\frac{1}{2}\frac{\muF_{\theta'}^2-(\muF_{\theta'}-2\imag\sigmaF_{\theta'}^2)^2}{\sigmaF_{\theta'}^2}\right)-\mu_f^2.
\end{align}